\documentclass{aa}  
\usepackage[dvipsnames]{xcolor}  
\usepackage{lineno} 

\usepackage{hyperref}
\usepackage{tabularx}
\usepackage{graphicx}
\usepackage{txfonts}
\usepackage{hyperref}
\usepackage{url}

\titlerunning{Internal magnetic field structures observed by PSP/WISPR in a filament-related coronal mass ejection}

\authorrunning{ G.M. Cappello et al.}

\begin{document}

   \title{Internal magnetic field structures observed by PSP/WISPR in a filament-related coronal mass ejection}

   \subtitle{}

   \author{G.M. Cappello
          \inst{1}, M. Temmer\inst{1}, A. Vourlidas\inst{2}, C. Braga \inst{2}, P.C. Liewer\inst{3}, J. Qiu\inst{4}, \\ G. Stenborg\inst{2}, A. Kouloumvakos\inst{2}, A.M. Veronig\inst{1,5}, V. Bothmer\inst{6}
          }

   \institute{Institute of Physics, University of Graz, Universitätsplatz 5, 8010 Graz, Austria\\
              \email{greta.cappello@uni-graz.at}
         \and
             The Johns Hopkins University Applied Physics Laboratory, 11101 Johns Hopkins Road, Laurel, MD 20723, USA.
          \and   
             Jet Propulsion Laboratory, California Institute of Technology, Pasadena, CA, 91109, USA
        \and
            Department of Physics, Montana State University, Bozeman, MT, 59717, USA
        \and
            Kanzelh\"ohe Observatory for Solar and Environmental Research, University of Graz, 9521 Treffen, Austria
        \and
            Institut für Astrophysik und Geophysik, Georg-August-Universität, Göttingen, Germany
             }             
   \date{Received February 14, 2024; accepted April 24, 2024}

\abstract
  {We investigated the coronal mass ejection (CME) related to an eruptive filament over the southwestern solar limb on December 8, 2022, at around 8 UT. We tracked localized density enhancements reflecting the magnetic structures using white-light data taken with the Wide-field Instrument for Solar PRobe (WISPR) aboard the Parker Solar Probe (PSP).}
   {We aim to investigate the 3D location, morphology and evolution of the internal magnetic fine structures of CMEs. Specifically, we focused on the physical origin of the features in the WISPR images, how the white-light structures evolve over time, and their relationship with the source region, filament, and the flux rope. 
   }
   {The fast tangential motion of the PSP spacecraft during its perihelion permits a single event to be viewed from multiple angles in short times relative to the event's evolution. Hence, three-dimensional information of selected CME features can be derived from this single spacecraft using triangulation techniques. }
   { We grouped small-scale structures with roughly similar speeds, longitude, and latitude into three distinct morphological groups. We found twisted magnetic field patterns close to the eastern leg of the CME that may be related to "horns" outlining the edges of the flux-rope cavity. We identified aligned thread-like bundles close to the western leg, and they may be related to confined density enhancements evolving during the filament eruption. High density blob-like features (magnetic islands) are widely spread in longitude ($\sim$40°) close to the flanks and the rear part of the CME. We also note that the large-scale outer envelope of the CME, seen clearly from 1 AU, was not well observed by PSP.}
   {We demonstrate that CME flux ropes, apart from the blobs, may comprise different morphological groups with a cluster behavior; the blobs instead span a wide range of longitudes. This finding may hint at either the three-dimensionality of the post-CME current sheet (CS) or the influence of the ambient corona in the evolutionary behavior of the CS. Importantly, we show that the global appearance of the CME can be very different in WISPR (0.11--0.16~AU) and the instruments near 1~AU because of the shorter line-of-sight integration of WISPR.}


   \maketitle
%

\section{Introduction}

Coronal mass ejections (CMEs) refer to the substantial expulsion of plasma and magnetic flux from the solar corona into interplanetary space. Their driver is a magnetic flux rope structure, which may not necessarily contain cold filamentary material \citep[e.g.,][]{Chen2011}, accelerated outward by magnetic reconnection through the Lorentz force \citep{Forbes2000,Vrsnak2001}. Using multiple viewpoints of CMEs through stereoscopic coronagraph images from the Solar TErrestrial RElations Observatory \citep[STEREO;][]{Kaiser_STEREO}, valuable insights have been gained about their three-dimensionality when observed from 1~AU. Typically, but not always, a CME morphology is observed that includes a faint front (a shock sheath if the CME is fast enough) followed by a bright loop-like leading edge and then a dim region corresponding to a low plasma-beta structure delineating the flux rope \citep{Vourlidas2013,TB22}. When interacting with the near-Earth environment, CMEs can drive a variety of space weather effects, such as geomagnetic storms, disruption of satellite communication systems, and ground-induced currents \citep[see reviews by e.g.,][]{Pulkkinen2007,Gopalswamy2022}. Hence, the evolution of CMEs from the Sun to Earth has received particular attention over the past years. Several recent reviews have given an overview of our current understanding of the physics of CME evolution, their interaction with the ambient solar wind, the state of the art in modeling and analysis methods, and the gaps that need to be filled in the future \citep[e.g.,][]{Green2018,Temmer2021,Zhang2021,Gopalswamy2022,Mishra2023,Temmer2023}. 

There are clear associations of CMEs with phenomena observed on the solar disk, such as flares, filament or prominence eruptions, coronal waves, and dimming regions, etc. \citep[see e.g.,][]{Chen2011}. We note that in this manuscript, we will use the terms "filament" and "prominence" interchangeably, since we study observations from different viewpoints and both terms refer to the same physical phenomenon. Specifically, filaments are considered to be proxies of sheared magnetic field arcades \citep{DeVore2000ApJ_filament_sheared_arcades}, and magnetic flux ropes (MFRs), which snake along and above the neutral line \citep[e.g.,][]{Moore2001,Amari2003ApJ_filament_fluxrope}. Statistical studies have found that more than 80\% of erupting filaments are associated with a CME \citep[e.g.,][]{Schmieder2013}. They are particularly well observed in chromospheric spectral lines, such as H$\alpha$, and cool EUV lines, such as HeI 304\AA~\citep[e.g.,][]{Parenti2014, Chen2014ApJ}. 

High-resolution observations have indicated that filaments are composed of a collection of thin threads situated above the photospheric magnetic neutral line \citep[e.g.,][]{Tandberg-Hanssen1995, Aulanier2002, Okamoto2007}. Horizontal threads may be associated with cool material suspended between the dips of a sheared arcade or twisted flux rope, while vertical threads are not yet well understood \citep[see e.g.,][]{Su2015ApJ, Guo2021ApJ}. \cite{Schmieder2014} and \cite{Ruan2018} have argued that vertical threads are caused by the accumulation of many small dips containing short threads.

Although small-scale CME structures have been studied in the past, the accessible scales depend on the resolution of the imaging instrument at hand. Every new generation of white-light instruments reveals ever-finer details about the internal structures of CMEs and their evolution \citep{Howard_2023}. Here, we investigate a variety of transient density enhancements in order to trace the magnetic fine structure of a filament-related CME observed by the Wide-Field Imager for Solar PRobe \citep[WISPR;][]{Vourlidas2016} on board the Parker Solar Probe \citep[PSP;][]{Fox2016}.

WISPR provides stunning views of the off-limb corona with a higher sensitivity compared to coronagraphic imagery from missions located near 1~AU, such as the Solar and Heliospheric Observatory \citep[SOHO; see][]{Brueckner_1995_lascoc2, Domingo_1995_soho} and STEREO \citep[see][]{Kaiser_STEREO,Howard_2008_secchi}. 
\cite{Illing_1983} first identified loop-like structures visible within the ejecta as bright coronal material in the form of a concave upward arch in images from the Solar Maximum Mission \citep[SMM; ][]{SMM1980}. These fine structures, namely the 'horns', were also observed by \cite{Vourlidas2013} in SOHO coronagraphic images. The structures highlight the magnetic flux rope (MFR) nature of the cavity. \cite{Webb1995} and \cite{Webb2003} performed statistical studies on SMM data of concave-outward regions, and they found that about half of them were followed by coaxial and bright rays, suggesting the formation of the current sheets. The latter, usually depicted as a 2D surface, may last several hours and extends to the outer corona. Post-CME blobs are usually related to the post-CME current sheet and plasmoid instabilities such as magnetic islands\citep[see, e.g.,][]{Ciaravella2008,Schanche16,Webb2016, Lee2020ApJ}. Many simulations have been performed to understand the nature of these blobs. For example, \cite{Riley_2007_bursty_reconnection} considered bursty reconnection in the current sheet, as modeled by a MHD simulation, and suggested the formation of blobs and their acceleration along the rays to be through the energy released during the reconnection process. Also, \cite{Poirier2023AA_model} modeled blobs by simulating an intermittent release of density structures generated by tearing-induced magnetic reconnection at the tip of helmet streamers and propagating in the slow solar wind. 

WISPR has already made close-up views of CMEs (e.g., for distances of $\sim$0.062~AU during Encounter 10, or E10). \citep{Howard2022} highlighted the presence of brightness inhomogeneities that appear as blobs and smaller MFR structures within the larger MFR. \cite{Patel_2023ApJ} studied an arc-shaped concave-up structure in a WISPR CME. In addition, PSP has traversed CMEs capturing both remote sensing data and in situ measurements of the plasma properties and the magnetic field. \citep{Romeo2023ApJ} reported magnetic field variations within a CME containing enhancements, small rotations, and turbulence. Other highlights of the PSP mission are summarized in \cite{Raouafi2023}.

Usually, 3D reconstructions of the global shape of a CME are performed based on multi-viewpoint images from spacecraft located at 1~AU. From these, a CME's propagation direction, angular width, and tilt are derived \citep{Thernisien2009,Kouloumvakos2022}. Another method is to obtain the CME direction and speed from elongation measurements of well-observed CME fronts using single spacecraft methods, such as Point-P and fixed-$\phi$ methods, applied to heliospheric images \citep[see e.g.,][]{kahler07,wood09,lugaz09}. However, a recent case study by \cite{Patel_2023ApJ} yielded unrealistic speed results, demonstrating the shortcomings of traditional approaches for deriving kinematics when imaging in proximity to a transient. 

A better approach to derive reliable kinematics and directions of coronal structures or CMEs is the geometric triangulation method developed specifically for WISPR by \cite{Liewer2019, Liewer2020}. It leverages the fast motion of the PSP spacecraft and uses observations of a single event from multiple angles across a short time frame to perform triangulation with only a single spacecraft. This allowed us to track and investigate individual small-scale structures, that are part of the global CME. The technique has already been successfully applied in PSP studies to obtain the characteristics of CMEs and other coronal structures \citep{Liewer2020, Liewer2021,Liewer2022}, as well as the deflection \citep{Braga2021} and deformation of CME frontal structures \citep{Braga2022}.  

In this study, we investigate the evolution and morphology of small-scale magnetic field structures observed inside a CME, and we try to understand their relation to the associated erupted filament imaged by EUV instruments from 1~AU.  Specifically, we are concerned with how these structures form and develop and how they related to the larger-scale MFR in which they are embedded. The proximity of WISPR to the Sun further allowed us to study the effect of projection and line-of-sight integration on the CME intensity and to investigate where the small-scale structures are located with respect to the global appearance of the CME.  

This paper is organized as follows. In Sect. ~\ref{Event}, we describe the event and observational data used in this study. Section ~\ref{method} presents the methods used to investigate the small-scale structures observed during the passage of the CME with observations from PSP and also covers the CME 3D global reconstruction. In Sect. ~\ref{results} we present our results for the morphology and 3D information, which are discussed and summarized in Sect. ~\ref{discussion}. Finally, Appendix~\ref{appendix} covers a more detailed description of the geometrical triangulation method applied to the PSP/WISPR observations.

\section{Event description and data sources}\label{Event}

\begin{figure*}[!hbt]
\centering
\includegraphics[width=1.\textwidth]{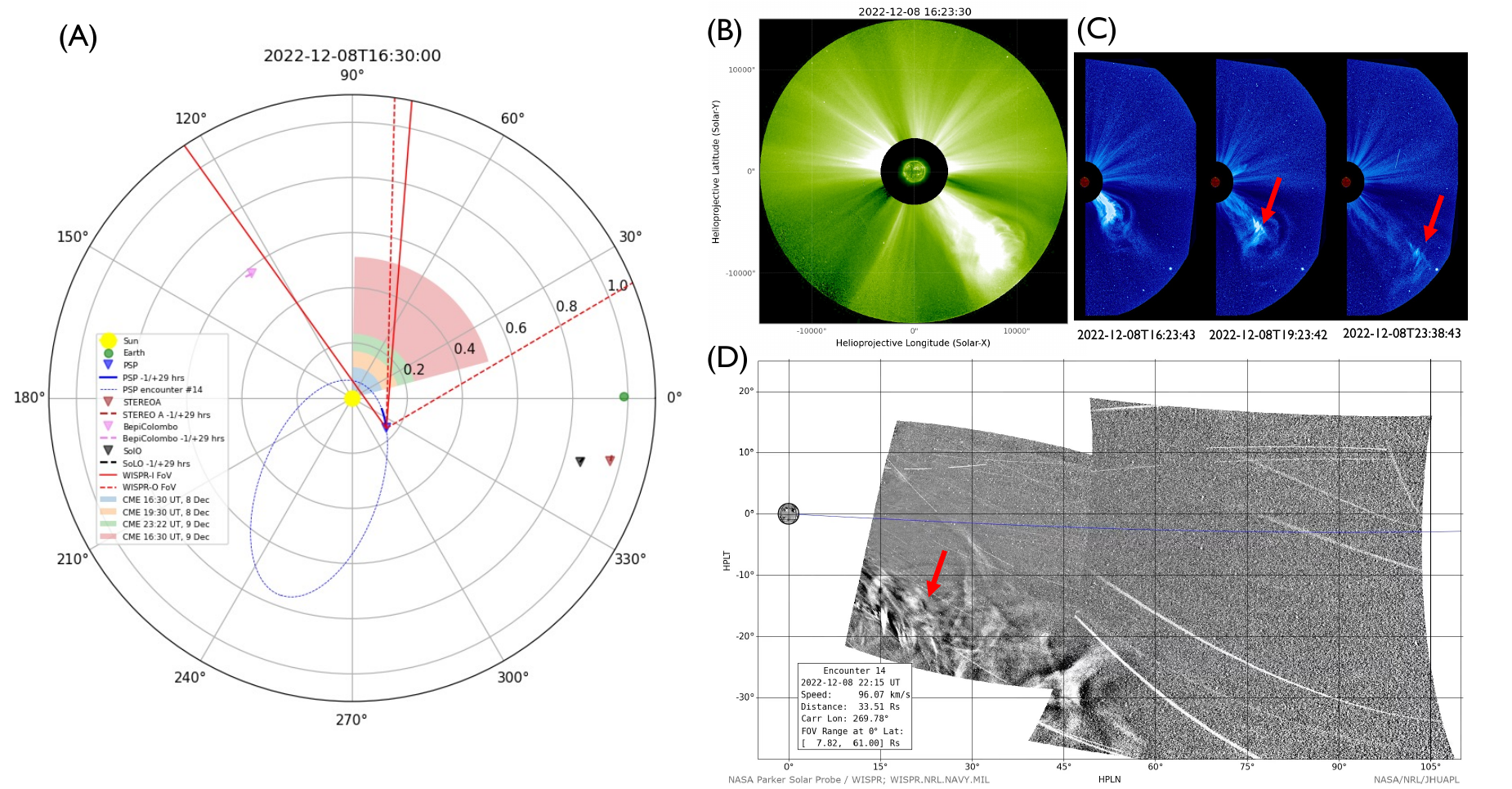}
\caption{Overview of the December 8, 2022 event showing the spacecraft constellation (panel A),  using an HCI coordinate system, together with white-light imaging data (with EUV inserts from SUVI in 304 \AA, and EUVI in 195 \AA) recorded from the various spacecraft vantage points. Panel B shows the vantage point from STA/COR2, panel C shows the vantage point from SOHO/LASCO/C3 and panel D, shows the vantage point from PSP/WISPR. The red arrows in panels (C) and (D) highlight the localized density enhancements, referring to the magnetic fine structures contained in the core of the CME (see Sect. ~\ref{eruption}). The spacecraft constellation shows PSP's orbit for E14 with the blue dashed line and the PSP trajectory during the event with the blue solid line. The PSP longitude changed $\sim24^\circ$ over the entire duration of the event (15:30 UT on December 8 to 23:30 UT on December 9). The FoV of WISPR-I (red solid line) and WISPR-O (red dashed line) are also shown together with the CME propagation direction and width, which was obtained using the GCS reconstruction results (see Sect. ~\ref{GCS}), at four different instants of time (16:30~UT [0.11 AU], 19:30~UT [0.17 AU], and 23:22~UT [0.23 AU] on December 8 and 16:30~UT [0.51 AU] on December 9). }
\label{overview}
\end {figure*}

We investigated the CME event that occurred on December 8, 2022,  and is associated with an eruptive filament over the southwestern limb of the Sun (from Earth view). The first CME front appeared in the LASCO/C2 field of view (FoV) at 04:12~UT and in C3 at 10:42~UT (at a position angle of 240°). According to the SOHO/LASCO CME catalog\footnote{\url{https://cdaw.gsfc.nasa.gov/CME_list/}}, the CME had a projected linear speed of $\approx$ 300 km/s and an angular width of 100°. 

For the event analysis, we used data from various viewpoints and instruments covering the white-light and EUV wavelength ranges. White-light data were obtained from the SOHO observatory \cite{Domingo_1995_soho} located at the Lagrangian point L1 and operating two white-light coronagraphs: LASCO/C2, with an field of view (FoV) of 1.5--6 R$\textsubscript{\(\odot\)}$, and LASCO/C3, with an FoV of 3.7--32 R$\textsubscript{\(\odot\)}$. The Sun-Earth Connection Coronal and Heliospheric Investigation \citep[SECCHI;][]{Howard_2008_secchi} aboard the STEREO-A (STA) spacecraft provides a different perspective, as STA observed the event from about 1~AU and 14° west of the Sun-Earth line. We used data from the SECCHI instrument suite, including the inner white-light coronagraph COR1 (FoV: 1.5--4 R$\textsubscript{\(\odot\)}$) and the outer COR2 (FoV: 2.5--15 R$\textsubscript{\(\odot\)}$), as well as white-light heliospheric imagers, HI-1 and HI-2, covering an FoV up to Earth location and beyond. 

Our main focus is the WISPR data, as they provide unprecedented remote sensing white-light images from a close-in orbit. WISPR comprises two telescopes. The inner one, WISPR-I, images an elongation of 13.5$^\circ$--53.0$^\circ$, while the outer, WISPR-O, covers 50.5$^\circ$--108.5$^\circ$. For more details on the WISPR instrument, we refer to \cite{Vourlidas2016}. 

Due to the PSP highly elliptical orbit, its heliocentric distance and its viewing angle changes drastically during its orbit around the Sun. Therefore, the WISPR FoVs and temporal cadences change accordingly. The event under study was observed during Encounter 14 (E14). We used observations from December 8, 2022, 15~UT to December 9, 2022, 23:30~UT. During that time, PSP moved from $-$41.9$^\circ$ (0.16 AU) to $-$18.2$^\circ$ (0.11 AU) in longitude (heliocentric distance) in the heliocentric Inertial (HCI) coordinate system. 

The time cadence increased from $\sim$30 min and $\sim$5 min as the spacecraft approached the Sun during the event. Furthermore, the event was captured in the "wave turbulence" mode (WT), where a subfield in WISPR-I is captured at 1-min cadence for several hours. The WT data from 08:03 UT to 20:53 UT on December 8, 2022, are available. No WISPR-O images were acquired during the WT mode, but WISPR-I full FoV images were available. The WT image data allowed us to follow the small-scale structures with less ambiguity and to better appreciate the complexity of the CME. We made use of the "LW"-processed WISPR images for the analysis (see the appendix in \cite{Howard2022}, for more details). The LW technique increases the contrast and facilitates the tracking of the features. 

To relate the features observed in white-light to the solar source region, we used the available EUV data in the wavelength ranges of 171~\AA~, 195~\AA~, and 304~\AA~ from the Extreme Ultraviolet Imager (EUVI) on board STA with an FoV up to 1.7R$_\odot$ \citep{Wuelser04}. As it was close to the solar limb, the location of the erupting promience allowed us to track the event back to the solar surface. For the Earth perspective, we used EUV data from the Solar Ultraviolet Imager \citep[SUVI;][]{SUVI_GOES_DARNEL} onboard the Geostationary Operational Environmental Satellite\citep[GOES-R;][]{Krimchansky_goes2004}, which has an FoV similar to EUVI. We note that we do not discuss the initiation and triggering of the filament eruption itself. 

The left panel of Fig. ~\ref{overview} provides an overview of the spacecraft configuration on December 8, 2022, and marks the CME trajectory at different times. The right panels in Fig. ~\ref{overview} show combined COR2/EUVI-195\AA~ (panel B), C3/SUVI-304\AA~ (panel C), and WISPR-I/O composite snapshots (panel D). The COR2 snapshot was obtained at 16:23~UT. The C3 images were taken at 16:23~UT, 19:23~UT, and 23:38~UT, and they reveal aligned fine structures (marked by a red arrow in panel C). The WISPR composite, taken at around 22:15~UT (or $\sim 22.22$ UT in Earth time), shows an incredible amount of detail within the CME, marked with a red arrow in panel D.

\section{Methods}\label{method}

Imaging from a single spacecraft cannot directly derive the 3D structure of a CME since it only captures a 2D projection in the plane of the image. Several techniques for CME 3D reconstructions, especially for their extended fronts, were developed during the STEREO era, and offered at least two different vantage points from where CMEs could be observed \citep[see, e.g.,][]{lugaz09, Temmer2009,Byrne2010,Mierla2010,rollett2016ApJ}. The most widely used method is the so-called graduated cylindrical shell (GCS) model \citep{Thernisien2009, Thernisien2011}. The parameters used to define the GCS model are the longitude and latitude at the apex of the CME; its height; the rate of expansion $\kappa$; the half angle $\alpha$, which represents the angle between the main axes and half of one leg; and the tilt with respect to the ecliptic plane. We applied the GCS model to stereoscopic image data from STA and SoHO, both located at about 1~AU (separation angle $\approx$ 14$^\circ$), and PSP. The software package PyThea \citep{Kouloumvakos2022} was used to perform the GCS reconstruction on the HI-1 data, while the package developed by \cite{Forstner21} was used for the reconstruction on COR2, C3, and WISPR data. 

As PSP moves quite fast (around 160~km/s near the E14 perihelion), the single spacecraft image data can be used to derive the 3D morphology and kinematics. The method is described in \cite{Liewer2020}, and details of its application in this study are given in Appendix \ref{appendix}. The method assumes that the structures move radially outward at a constant velocity. Another  requirement is that the spacecraft angular position varies sufficiently, for which we take a minimum of  4$^\circ$. Using Eqs.~(\ref{eq:geometic1})-(\ref{eq:geometic2}), we derived the velocity, position, latitude, and longitude of a tracked small-scale structure. To solve the equations, we followed the approaches of both \cite{Liewer2020} and \cite{Braga2021}. Uncertainties were estimated by tracking each identified feature at least three times. 

\section{Results}\label{results}
 
\subsection{Small-scale structure morphology}

\begin{figure*}[!hbt]
\centering
\includegraphics[width = 0.7\textwidth]{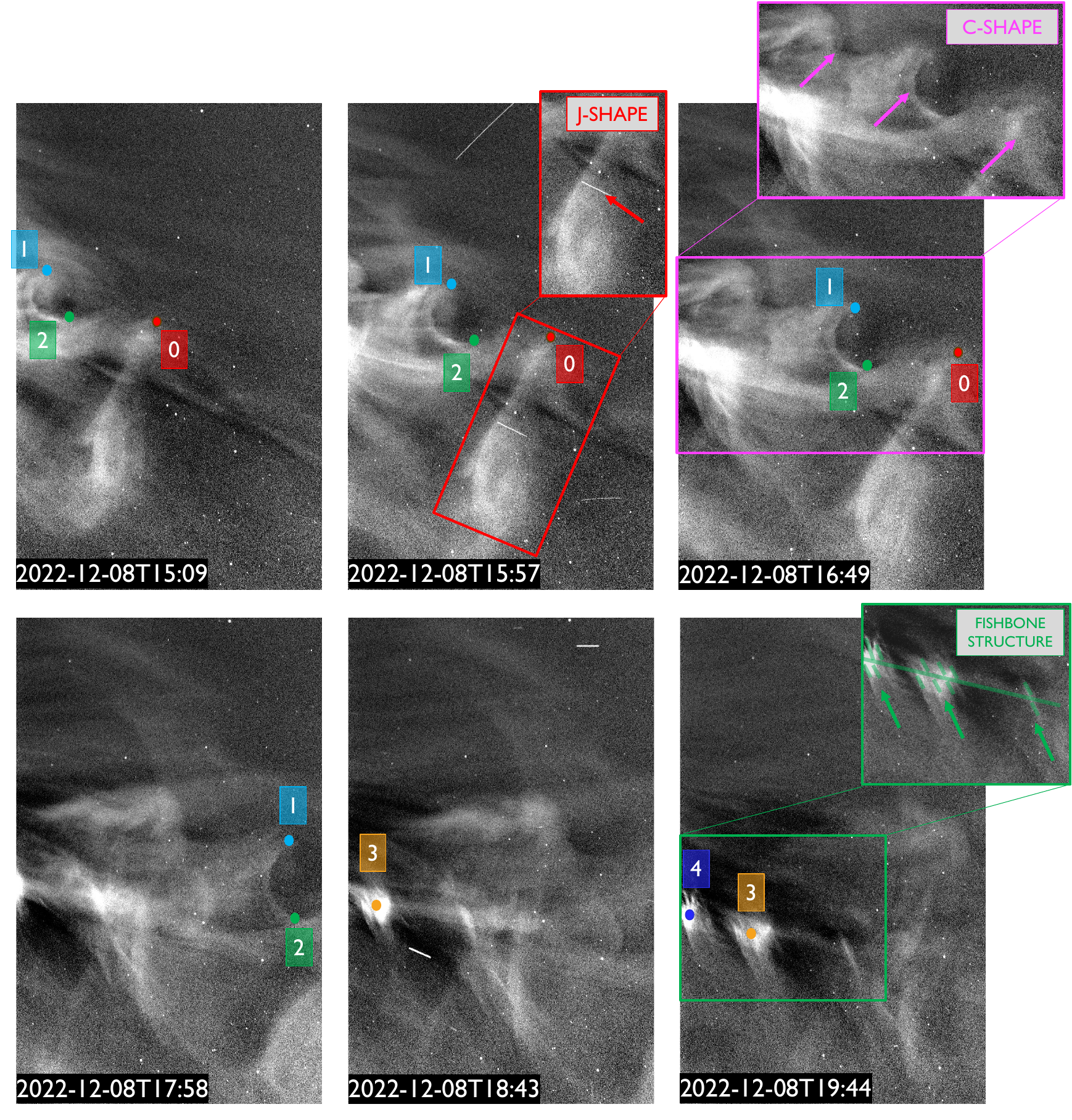}
\caption {Selected LW-processed snapshots of the WISPR-I WT data set obtained on December 8, 2022, between 15 UT and 20 UT showing the tracked features (labeled 0 through 4). The tracked features belong to structures resembling J-shapes (0), C-shapes (1, 2), and fish-bone-like shapes revealing an inclined thread-like pattern (3, 4). The x-axis shows the helioprojective longitude; the y-axis shows the helioprojective latitude. The animation corresponding to this figure is available online in the file movie1.mov.
}
\label{high_cadence_mosaic}
\end{figure*}

\begin{figure*}[!hbt]
\centering
\includegraphics[scale = 0.33]{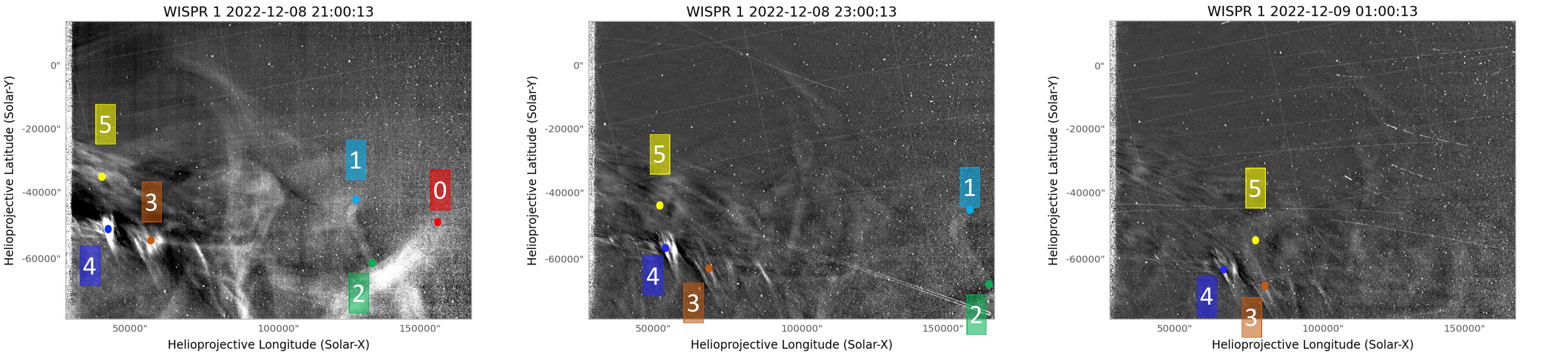}
\caption {WISPR-I images from December 8 20:30 UT to December 9 18:30 UT and tracked features (0--9). The information from these images is complementary to that of Fig \ref{high_cadence_mosaic}. }
\label{wl_mosaic}
\end {figure*}

\begin{figure*}[!hbt]
\centering
\includegraphics[scale = 0.4]{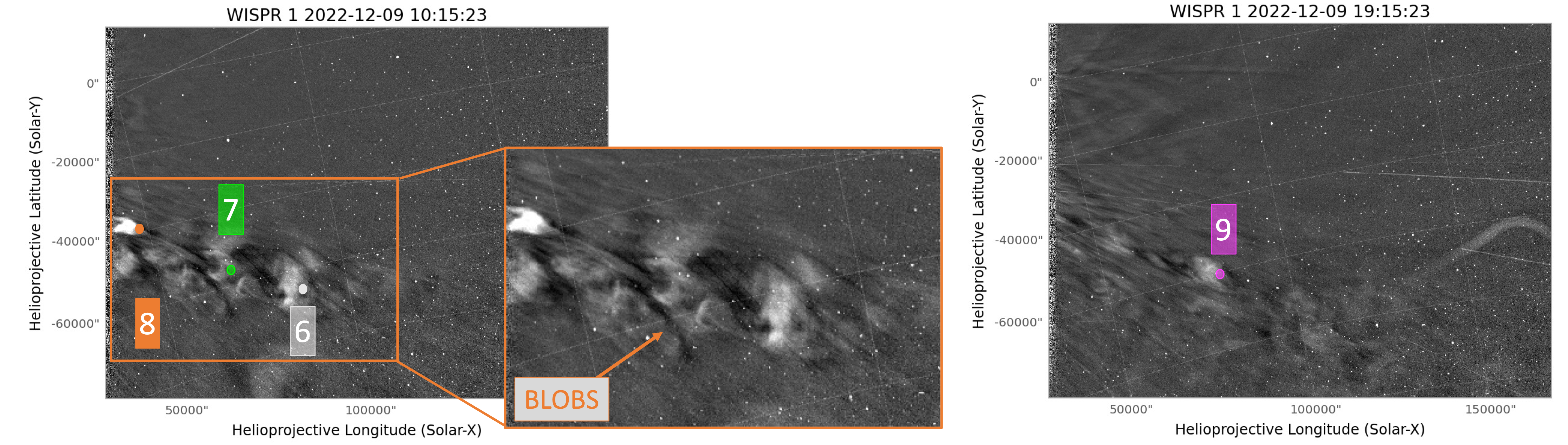}
\caption {Features 6 to 9 (blob-like) propagate in the wake of the CME. The information from these images is complementary to that from Figs. \ref{high_cadence_mosaic}-\ref{wl_mosaic}.}
\label{BLOBS_mosaic}
\end {figure*}

In Figs.~\ref{high_cadence_mosaic}-~\ref{BLOBS_mosaic}, we display snapshots of the selected small-scale structures, which we observed as the CME evolved within the FoV of WISPR-I and WISPR-O. All the tracked small-scale structures have been carefully selected, considering a significant variation of the HCI longitude of PSP, together with the sufficient visibility/contrast that they need to have over time (see Appendix \ref{appendix}).   
In total we tracked ten different features over a period of about 32 hours (labeled 0 through 9, in the chronological order in which they entered the WISPR-I FoV). In particular, Fig.~\ref{high_cadence_mosaic} shows the tracked features 0--4 of the CME in a section of the FoV of WISPR-I, when the instrument was in the WT mode (see Sect. ~\ref{Event}). The tracked features clearly outline curved geometries, which might be related to twisted magnetic field structures. A J-shaped structure, that resembles half of a U-shape, is clearly visible; therefore, we performed a track at its upper edge (feature 0), that stays longer in both FoVs of WISPR-I and WISPR-O with respect to the lower edge. Multiple C-shaped structures were also observed, with track features 1 and 2 located on the upper and lower edges of the "C" shape. Also present is a well-defined thread-like structure (features 3 and 4), which is typical for filaments. 

\begin{figure}[!hbt]
\centering
\includegraphics[width = 0.5\textwidth]{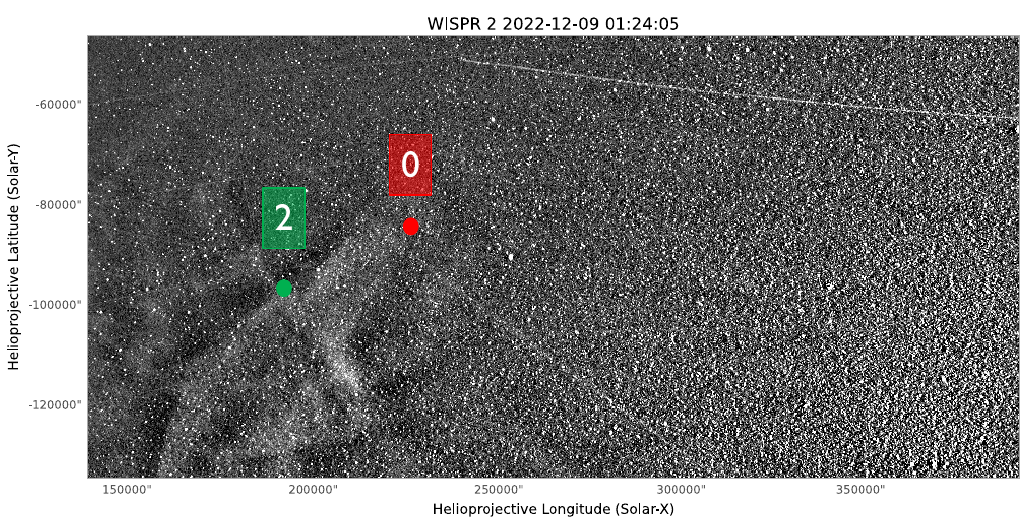}
\caption {WISPR-O images from December 8 23:39 UT and December 9 01:39 UT with tracked features 0 and 2.}
\label{track_outer_wispr}
\end {figure}

Figure \ref{wl_mosaic} shows the small-scale structures captured by WISPR-I in synoptic mode. An interesting item to note is another C-shaped feature, namely feature 5. As the synoptic mode covers a wider FoV, the left panel of Fig. \ref{wl_mosaic} reveals the total of features 0--5. Figure \ref{BLOBS_mosaic} shows a later time sequence on December 9, 2022, and the aftermath of the CME. Here, a series of blobs (features 6, 7, 8, 9) is visible, propagating in the wake of the CME. As time progresses, we noticed here how the C-shaped structure at the front expands over time and changes its shape (compared to Fig.~\ref{high_cadence_mosaic}). Because the CME is directed southward, many features tracked in WISPR-I do not reach WISPR-O. In addition, the features become fainter as they move to the outer instrument. As a consequence, only features 0 and 2 can be clearly tracked in the FoV of WISPR-O, as shown in Fig. \ref{track_outer_wispr}. All other features were tracked using WISPR-I.

\subsection{Small-scale structure 3D single-spacecraft reconstruction}
\label{Small-scale3D}
Over the duration of the event from 15:30 UT on December 8 to 23:30 UT on December 9, PSP longitude varied by about $\sim$22$^\circ$ (HCI; cf.\, left panel of Fig. ~\ref{overview}). Each of the ten features was tracked multiple times with at least three repetitions and the methods by \cite{Braga2021} and by \citep{Liewer2020} were applied to derive their 3D trajectories. Appendix~\ref{appendix} covers all related plots for deriving the results using the two different approaches. The results are summarized for both methods in Table \ref{results_table}, which gives the HCI longitude and latitude as well as the initial speed and distance from the Sun for each feature. Additionally, we report the angular distance, $\Delta\phi_1$, PSP traveled in longitude (HCI) during the tracking of that specific feature; the angular distance, $\phi_2-\phi_1$, between the feature and PSP; and finally the morphological group.

Within the uncertainties, we obtained for the tracked features comparable results for all the parameters derived from the two different methods. All the results shown are in the HCI coordinate system. In general, all features propagate within a longitudinal range from 17$^\circ$ to 76$^\circ$ and a latitudinal range from $-$23$^\circ$ to $-$45$^\circ$. Inspecting the results in Table \ref{results_table}, we found that features 0, 1, and 2 propagate along 17$^\circ$ to 26$^\circ$ in longitude and between $-$23$^\circ$ and $-$30$^\circ$ in latitude, with average speeds of $\sim$480 km/s. Feature 5 propagates in a similar direction to features 0--2 (longitude from 31$^\circ$ to 38$^\circ$ and latitude from $-$36$^\circ$ to $-$37$^\circ$), though likely with a lower speed ($\sim$450 km/s). Features 3 and 4 propagate along a clearly different direction (longitude between 69$^\circ$ and 76$^\circ$, latitude from $-$43$^\circ$ to $-$45$^\circ$) with speeds between 300--400 km/s. The blob-like features 6--9 propagate at similar speeds ($\sim$300-400 km/s) and latitudes (from $-$31$^\circ$ to $-$39$^\circ$) but are distributed over a large longitudinal range between 24$^\circ$ and 66$^\circ$.

Based on their similarity in morphology and dynamics, the tracked features can be divided into three groups. Group 1 (features: 0, 1, 2, and 5) most likely refers to twisted magnetic field structures with C- and J-shapes. Group 2 (features: 3 and 4) refers to thread-like patterns, and group 3 (features, 6, 7, 8, and 9) refers to blobs propagating for several hours in the rear part of the CME. On average, we observed a decreasing trend in speed from the fastest group (group 1) to the slowest (group 3). While each feature of group 1 and 2 shows a clustered behavior, that is, they reveal a roughly similar speed, longitude, and latitude among each group, the blobs (group 3) span a wide range of longitudes ($\sim$ 40$^\circ$).

\begin{table*}
\caption{\label{results_table} Results in HCI obtained by applying the methods discussed in Sect.~\ref{method} to the features labeled in Figs. \ref{high_cadence_mosaic}-\ref{wl_mosaic}.}
\centering  
\begin{tabular}{ |p{0.2cm}||p{4cm}|p{0.6cm}|p{1.5cm}|p{1cm}|p{1.5cm}|p{2cm}|p{1.5cm}| p{1cm}|}
 \hline
 \multicolumn{8}{|c|}{Results of the feature tracking: \cite{Braga2021} (blue), \cite{Liewer2020} (red)} \\
 \hline
 ID & $t_{0}$ & $\Delta\phi_1$  (deg) &$\phi_2$  (deg) & $\phi_2$-$\phi_1$  (deg) & $\delta_2$  (deg) & $v_{0}$ (km/s)& $r_{0}$ (R$\textsubscript{\(\odot\)}$) & group\\
 \hline
 
 0& 2022-12-08T15:00:45.635& 6.2 & \textcolor{blue}{$22\pm 3 $} \newline  \textcolor{red}{$ 19 \pm 4 $} & \textcolor{blue}{64} \newline \textcolor{red}{61} & \textcolor{blue}{$-25 \pm 1$} \newline \textcolor{red}{$ -26 \pm 2 $} & \textcolor{blue}{$ 500 \pm 40 $} \newline \textcolor{red}{$ 464 \pm 62$} & \textcolor{blue}{$ 12.0 \pm 1.0$} \newline \textcolor{red}{$ 11.0 \pm 1.0$} & 1 \\

 \hline

 1 &  2022-12-08T15:00:45.635& 4.3 & \textcolor{blue}{$ 26\pm  2$} \newline \textcolor{red}{$ 32 \pm 9 $} & \textcolor{blue}{66} \newline \textcolor{red}{74} & \textcolor{blue}{$-23 \pm 1 $} \newline \textcolor{red}{$ -25\pm 3$} & \textcolor{blue}{$ 460 \pm 50 $} \newline \textcolor{red}{$ 492 \pm 106$} & \textcolor{blue}{$ 10.0\pm 1.0$} \newline \textcolor{red}{$ 10.0 \pm 2.0$} &1 \\
 
 \hline

  2 &  2022-12-08T15:00:45.635& 7.4 & \textcolor{blue}{$ 26 \pm 8 $} \newline \textcolor{red}{$ 17 \pm 8 $} & \textcolor{blue}{67} \newline \textcolor{red}{59} &\textcolor{blue}{$-30 \pm  3$} \newline \textcolor{red}{$ -29 \pm 3 $} & \textcolor{blue}{$ 486 \pm 80 $} \newline \textcolor{red}{$ 395 \pm 54 $} & \textcolor{blue}{$ 11 \pm 1.0 $} \newline \textcolor{red}{$ 11 \pm 1.0 $} & 1\\

  \hline

   3  &  2022-12-08T18:30:45.637& 5.0 & \textcolor{blue}{$ 71 \pm 11 $} \newline\textcolor{red}{$ 69 \pm 10 $} & \textcolor{blue}{111} \newline \textcolor{red}{109} & \textcolor{blue}{$-43 \pm 2 $} \newline \textcolor{red}{$ -43 \pm 1 $} & \textcolor{blue}{$ 383\pm 100 $} \newline \textcolor{red}{$ 386 \pm 50$} & \textcolor{blue}{$ 14.0\pm 1.0$} \newline \textcolor{red}{$ 13.0 \pm 0.4$} & 2 \\
  \hline
  
  4 & 2022-12-08T19:30:45.638& 6.3 & \textcolor{blue}{$ 76 \pm  5$} \newline \textcolor{red}{$ 70 \pm 10 $} &\textcolor{blue}{116} \newline \textcolor{red}{110} & \textcolor{blue}{$ -45\pm  1 $} \newline \textcolor{red}{$ -45 \pm 1 $} & \textcolor{blue}{$ 350 \pm 50 $} \newline \textcolor{red}{$ 293 \pm 57 $} & \textcolor{blue}{$ 13.0 \pm 1$} \newline \textcolor{red}{$ 13.0 \pm 0.2$} & 2\\

  \hline
  5 & 2022-12-08T21:30:48.797 & 7.2 & \textcolor{blue}{$ 31 \pm 2 $} \newline \textcolor{red}{$ 38 \pm 8 $} & \textcolor{blue}{70} \newline \textcolor{red}{77} &\textcolor{blue}{$-36 \pm 1$} \newline \textcolor{red}{$ -37 \pm 2 $} & \textcolor{blue}{$426 \pm 50 $} \newline \textcolor{red}{$ 467 \pm 47$} & \textcolor{blue}{$ 9.0\pm 0.2$} \newline \textcolor{red}{$ 9.0 \pm 0.6$} & 1 \\

 \hline

  6 &  2022-12-09T04:38:21.002 & 3.9 & \textcolor{blue}{$ 58 \pm 4 $} \newline \textcolor{red}{$ 53 \pm 5 $} & \textcolor{blue}{92} \newline \textcolor{red}{86} &\textcolor{blue}{$ -39\pm 1 $} \newline \textcolor{red}{$ -39\pm 1 $} & \textcolor{blue}{$ 316 \pm 25 $} \newline \textcolor{red}{$ 296 \pm 20$} & \textcolor{blue}{$ 9.0\pm 1.0$} \newline \textcolor{red}{$ 8.0 \pm 0.1$} & 3  \\

 \hline

  7 &  2022-12-09T07:00:41.103 & 5.4 & \textcolor{blue}{$43 \pm 2 $} \newline \textcolor{red}{$ 45 \pm 6$} & \textcolor{blue}{76} \newline \textcolor{red}{78} &\textcolor{blue}{$ -35 \pm 1$} \newline \textcolor{red}{$-35 \pm 4 $} & \textcolor{blue}{$ 400 \pm 10 $} \newline \textcolor{red}{$ 391 \pm 20 $} & \textcolor{blue}{$9 \pm 0.1 $} \newline \textcolor{red}{$ 9.0 \pm 0.1$} & 3 \\

 \hline

  8 &  2022-12-09T10:00:41.107 & 5.4 & \textcolor{blue}{$ 24 \pm 6 $} \newline \textcolor{red}{$ 38 \pm 6 $} & \textcolor{blue}{55} \newline \textcolor{red}{69} &\textcolor{blue}{$ -31\pm 2 $} \newline \textcolor{red}{$ -35\pm 1 $} & \textcolor{blue}{$ 300 \pm  50$} \newline \textcolor{red}{$  351 \pm 20$} & \textcolor{blue}{$ 9 \pm 1.0$} \newline \textcolor{red}{$ 8.0 \pm 0.2$} & 3 \\

 \hline
  9 &  2022-12-09T16:30:48.811 & 7.1 & \textcolor{blue}{$ 64 \pm 8 $} \newline \textcolor{red}{$ 66 \pm 4 $} & \textcolor{blue}{89} \newline \textcolor{red}{91} &\textcolor{blue}{$ -37\pm 1 $} \newline \textcolor{red}{$ -38 \pm 1 $} & \textcolor{blue}{$ 416 \pm 50 $} \newline \textcolor{red}{$ 425 \pm 22$} & \textcolor{blue}{$ 8.0\pm 1.0$} \newline \textcolor{red}{$ 8.0 \pm 0.4$} & 3\\

 \hline
 
 \hline
\end{tabular}
\newline
\tablefoot{The label of the feature, is indicated by its ID; $t_0$ is the initial time at which each feature was tracked; $\Delta\phi_1$ is the variation in the longitude of the PSP spacecraft during the tracking of the specific feature; $\phi_2$ and $\delta_2$ are the derived longitude and latitude, respectively; and $v_0$ and $r_0$ are the initial velocity and position of the feature. The last column represents the group in which the different features belong (see Sect.~\ref{Small-scale3D}). In blue we show the results obtained with the fitting method of \cite{Braga2021}, while in red we show the results from the fitting method of \cite{Liewer2020} (see Appendix \ref{appendix} for more details).}
\end{table*}

\subsection{Global GCS 3D reconstruction}\label{GCS}

\begin{figure}[ht!]
\centering
\includegraphics[width = 0.5\textwidth]{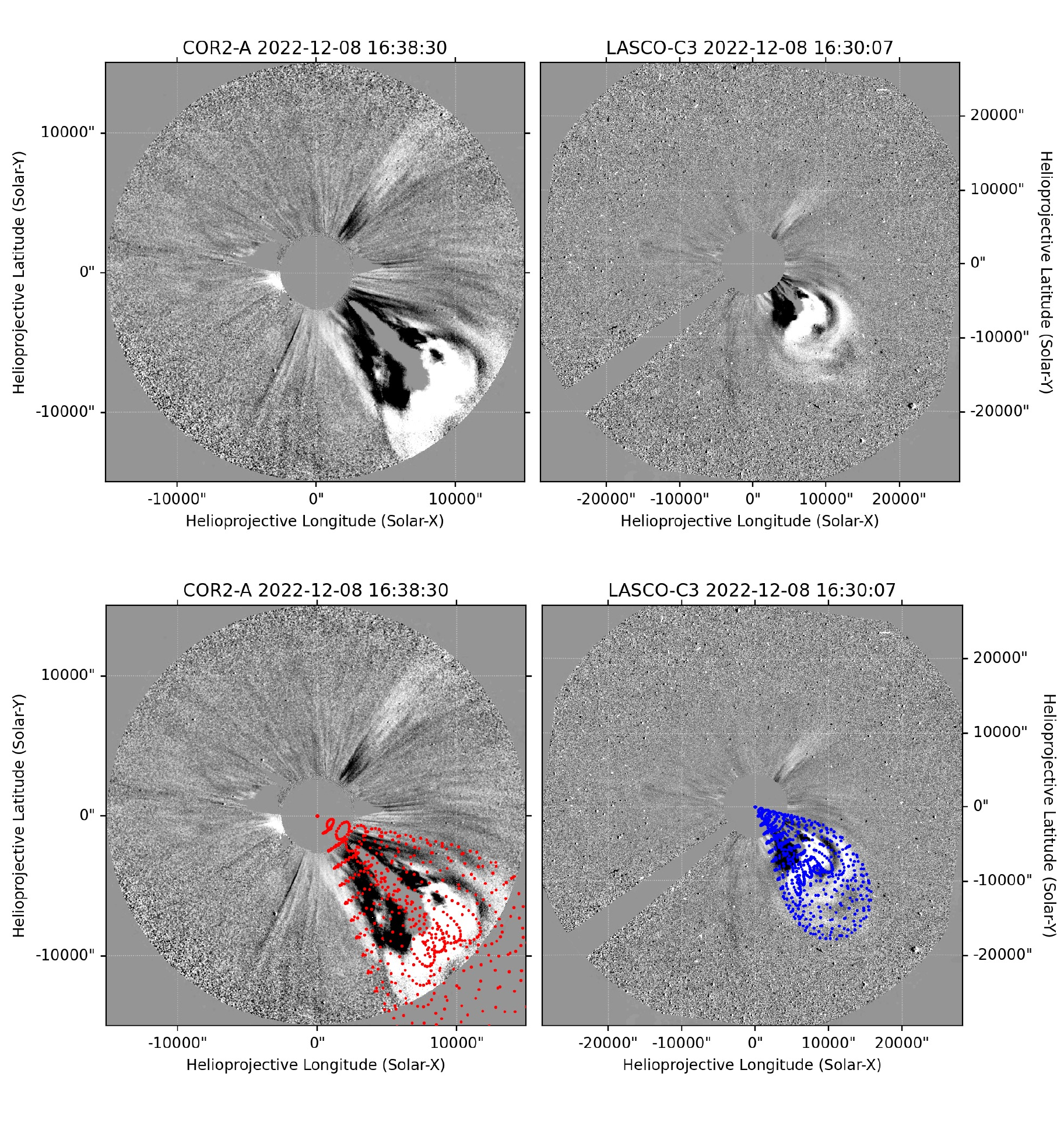}
\caption{GCS reconstruction of the running difference coronagraph images of COR2 and C3 at $\sim$16:30 UT. 
}
\label{gcs_coronographs}
\end {figure}

Comparison to the 3D reconstruction of the CME volume allowed us to interpret in more detail where the tracked small-scale structures are located with respect to the global appearance of the CME. The free parameters of the GCS model (the apex height of the idealized flux rope, the angular width, and aspect ratio) were fitted so as to match the global structure of the CME in stereoscopic images from C3 and COR2. Figure \ref{gcs_coronographs} shows the GCS reconstruction at $\sim$16:30 UT on December 8, 2022. The GCS fit indicates a propagation direction for the apex roughly along 52° in longitude and $-$42° in latitude (HCI coordinates) with uncertainties in the range of about 10° for latitude and longitude, respectively \citep[for detailed estimates on the accuracy of the GCS method we refer to][]{Verbeke2023}. The GCS geometry is described by the results $\alpha$=22°, $\kappa$=0.26 rad, and a tilt of $-$52$^\circ$. From this, we calculated the angular width face-on ($\omega_{\rm FO}$), which describes the lateral extension of the CME, by $\omega_{\rm FO}$=2($\alpha$+$\delta$) with $\delta=\sin^{-1}\kappa$, as well as the edge-on angular width $\omega_{EO}=2\delta$ \citep[see Table 1 in][]{Thernisien2011}. According to the GCS modelled flux rope, the CME has a half width, $\omega_{\rm FO}$, of $\sim$37° and an angular width at the edge $\omega_{EO}$ of $\sim30^\circ$. Therefore, the CME spans from about 15 to 89 in HCI longitude and from $-$12 to $-$72 in latitude. 

\begin{figure*}[ht!]
\centering
\includegraphics[scale = 0.30]{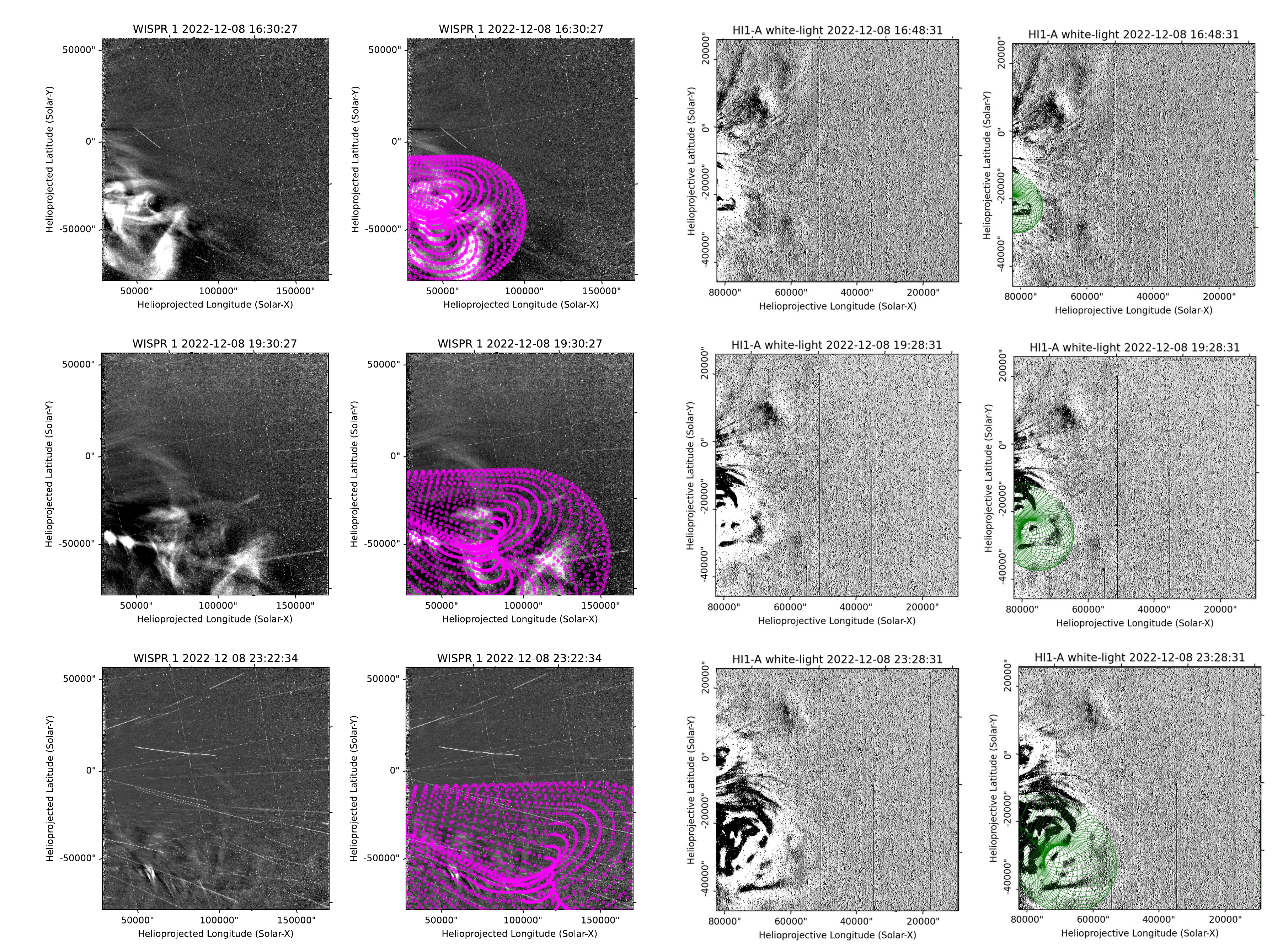}
\caption{GCS reconstruction of the heliospheric images of HI-1 and WISPR 1 at three instants of time.}
\label{gcs_hi}
\end {figure*}

Next, we expand the GCS fit to the inner Heliosphere by combining HI-1 and WISPR-I data.\footnote{We note that a $\sim$7 minute time delay between observations from 1~AU (STA) and from 0.11-0.16~AU (PSP) should be considered to account for the light travel time difference between PSP and the spacecraft at 1~AU.} For that we used the GCS results derived from the stereoscopic image data combining C3 and COR2 at $\sim$ 16:30 UT, kept all other parameters constant, and simply changed the height over time to match the CME front as observed in the heliospheric image data pairs. Figure \ref{gcs_hi} shows the GCS model applied to images from HI-1 and WISPR-I at $\sim$17:00 UT ($H$\,=\,24\,$R_\odot$), $\sim$19:30 UT ($H$\,=\,36\,R$_\odot$), and $\sim$23:22 UT ($H$\,=\,50\,$R_\odot$). With this approach, we derive a 3D CME speed of about 650 km/s. 
We observed distinct differences in the appearance of the CME between WISPR and HI-1. The WISPR images contain more small-scale structures, but the broad outer envelope is missing. The latter is clearly visible in both COR2 and HI-1.


\begin{figure*}[!hbt]
\centering
\includegraphics[width=1.0\textwidth]{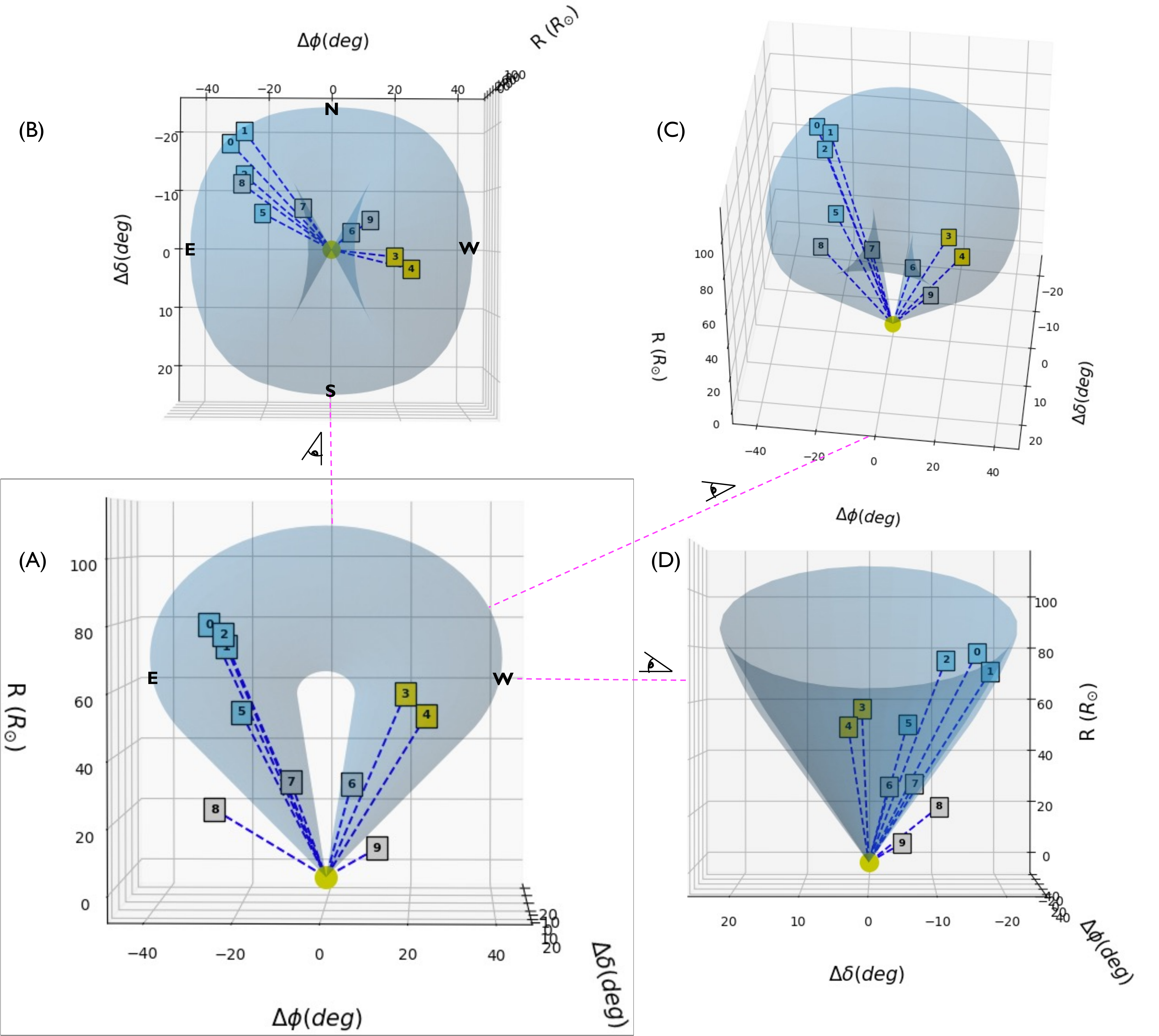}
\caption {Comparison of the positions of the local magnetic structures summarized in Table \ref{results_table} together with the global CME flux rope modeled by GCS. All panels show 3D plots in a $\Delta\phi$, $\Delta\delta$, $R$ plane at 16:30 UT on December 9, 2022. The relative variation in longitude and latitude with respect to the central axis of the GCS are represented by $\Delta\phi$ and $\Delta\delta$, while R is the height. The letter W indicates the west, E is east, S is south and N is north. The eye represents the point of view of the observer. The colors of the boxes indicate the different families to which the features 0 to 9 belong: 1 (azure), 2 (yellow), and 3 (gray). The plot shows the following: (A) the frontal view of the GCS modeled flux rope and the relative location of the features with respect to its central axis; (B) the extension of the flux rope from a top view; (C) same as plot B but from a west-top view, (D) same as plot B but from the western leg.}
\label{3d_view}
\end{figure*}

To visualize the location of the different features in 3D and relate the results in Table \ref{results_table} with those of GCS, we calculated the position of each small-scale structure at 16:30 UT on December 9, 2022 ($t_0$ of feature 9). Using the derived 3D speed, we extrapolated the height of the GCS apex accordingly to 110 R$_\odot$ for the same time. Figure \ref{3d_view} shows the resulting 3D distribution of the features with respect to the central axis of the GCS located at $\Delta\phi = 0$ deg, $\Delta\delta = 0$ deg and spanning 0--110 R$_\odot$ in height. The tracked small-scale structures are shown in boxes whose color depends on the morphological group discussed in Sect.~ \ref{Small-scale3D}. Features 0--7 were all found to lie within the GCS reconstructed volume and can therefore be safely interpreted as being CME internal small-scale structures. Comparing the location of the different groups, as listed in Table \ref{results_table}, to the central axis of the GCS, we found that features 0, 1, 2, and 5 are located in the eastern leg of the CME, while features 3 and 4 are part of the western leg. 

The blobs, covering tracked features 6--7, seem to be located at the rear part of the CME and close to the inner regions of each flank. We note that features 8 and 9 appear to lie outside of the legs modeled by GCS by about 10°, which is in the range of uncertainties of the GCS parameters \citep{Verbeke2023}. We note that we did not include the tilt in the visualization. 

\subsection{Relation to the erupting filament}
\label{eruption}

\begin{figure*}
\centering
\includegraphics[width = \textwidth]{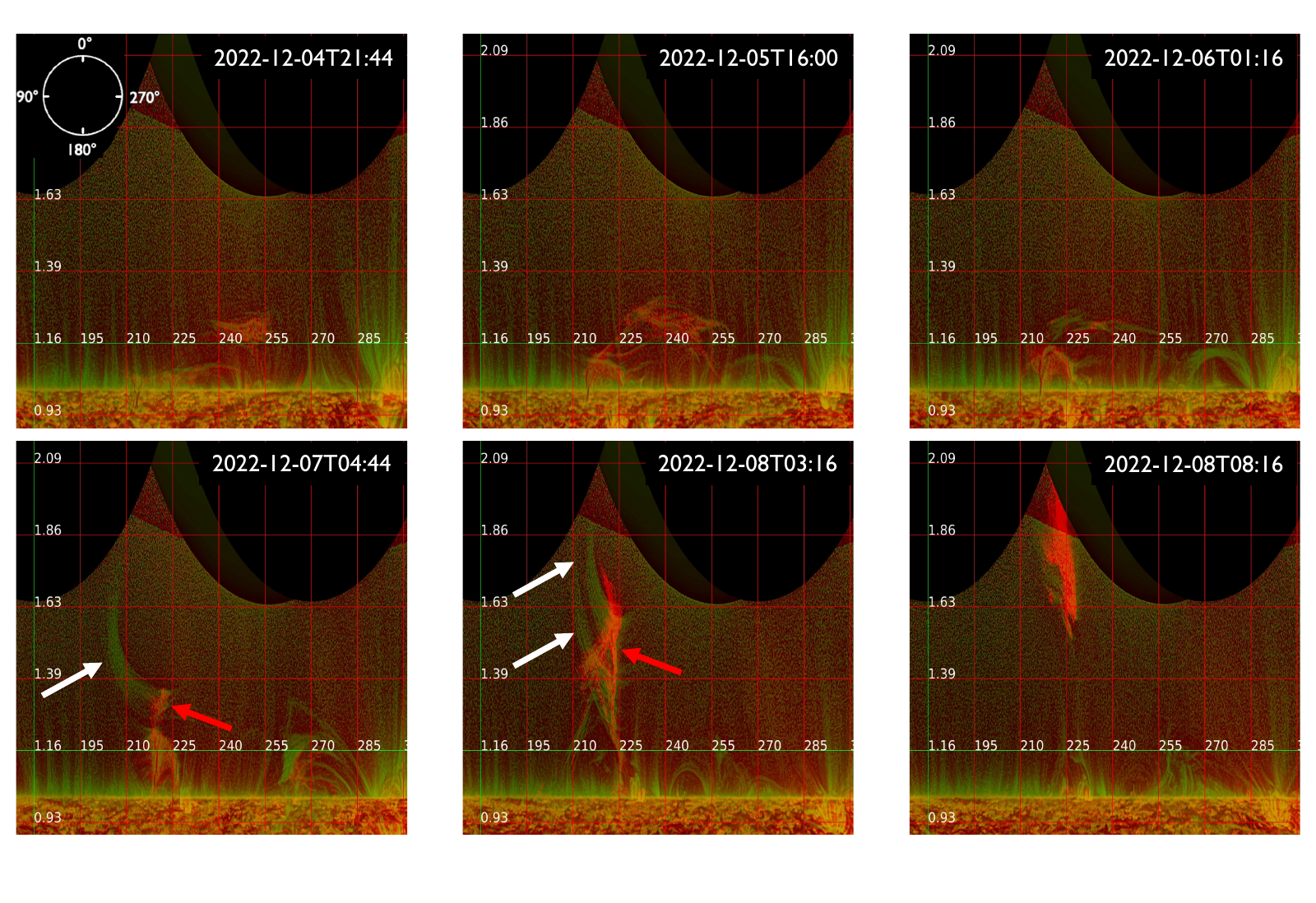}
\caption{Polar view of the solar prominence and its eruption in composite images of 304\AA~ and 171\AA~ observed by SUVI from December 4 to 8. In the y-axis, the height is in solar radii, while in the x-axis, the position angle is given. For orientation, in the top-left corner, a sketch of the PA is given, where 0° corresponds to the Solar North. The image was produced by JHelioviewer \citep{Jhelioviewer_2017}. The animation corresponding to this figure is available online in the file movie2.mov.}
\label{polar_euv_304}
\end{figure*}

To better understand the white-light morphology of the CME and its small-scale structures, we examined the structure and dynamics of the CME eruption from its coronal origin observed in EUV. This event can be clearly associated with a filament eruption from the southwestern limb of the Sun. Figure \ref{polar_euv_304} shows a polar projection view of the composite SUVI images in the 304\AA~ and the 171\AA~ filters. We note that the filament began crossing the western solar limb on December 4, and we can follow its height over the surface. The eruption and first appearance of the CME in COR1 occurs on December 8, 2022, at $\sim$1~UT. In the lower corona ($<$ 5 R$_\odot$), the filament rises at a speed of 11$\pm$4\,km/s, reaching about 310$\pm$18\,km/s (kinematics are not shown) after the eruption ($>$ 9 R$_\odot$). As the filament lifts off, a hot (green, 195 \AA~) "horn-shaped" structure (Fig. ~\ref{polar_euv_304}, white arrow) offers evidence of an MFR, and it is followed by cool (red, 304\AA) radial structures highly reminiscent of filament fine structures.

\begin{figure*}
\centering
\includegraphics[width = \textwidth]{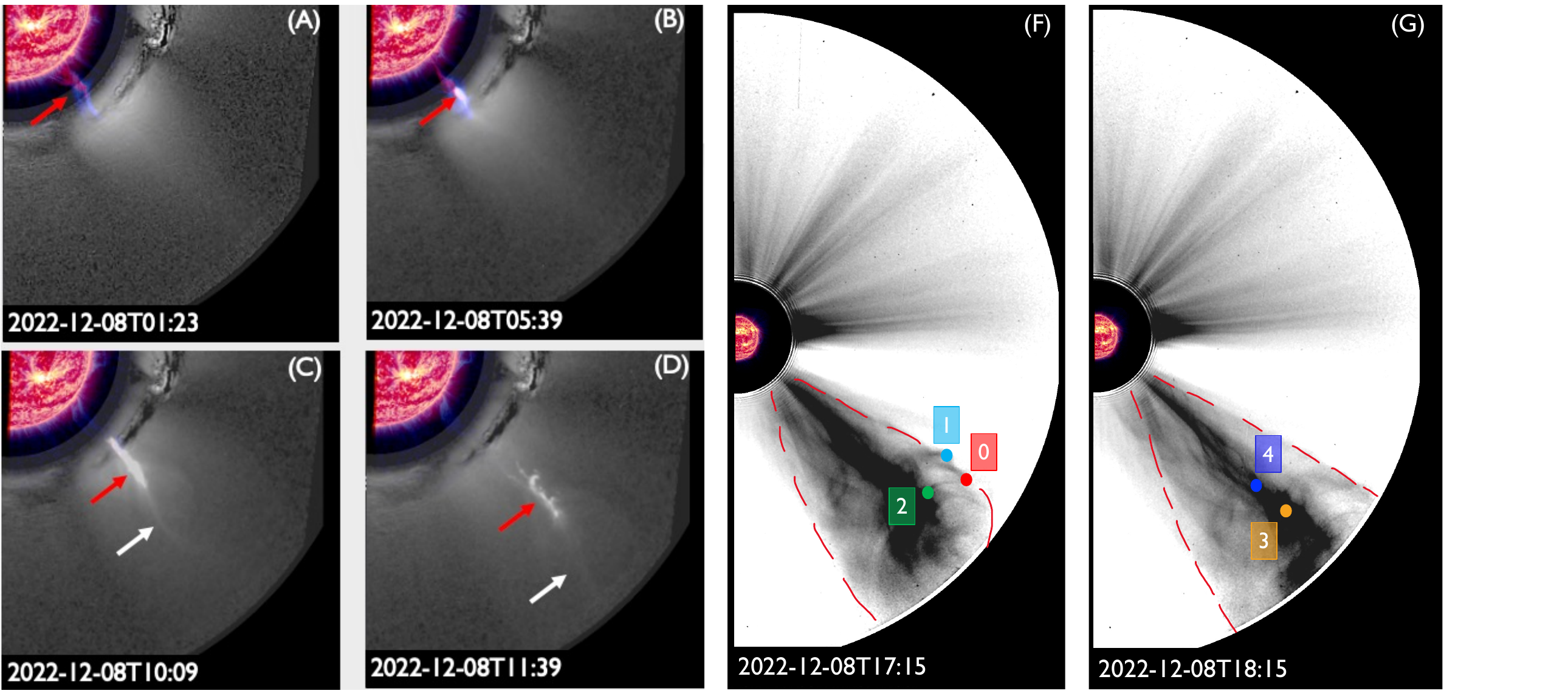}
    \caption{Evolution of the CME between the EUV instruments and coronagraphs. Panels (A)-(D) show composite EUVI 171 (blue)/304\AA (red) images and STA/COR1 from December 8 between 01:23 and 11:39~UT. The white arrow indicates the hot horn shape, and the red arrow points to the cool filament material. Panels (F) and (G) show the features tracked in WISPR images with COR2 identifications (see Fig. ~\ref{high_cadence_mosaic}).  
    }
\label{euv_cor1_counterpart}
\end{figure*}

In Fig. \ref{euv_cor1_counterpart}, we show combined EUVI and coronagraph images where one can clearly follow the evolution of these features from the EUV to the white-light images. Specifically, panels (A)--(D) show composite EUVI 171/304\AA~images  with COR1 data on December 8 between 01:23 and 11:39 UT. The white arrow indicates the hot horn shape, while the red arrow points to the cool filament material. In panels (E) and (F), we show COR2 images one hour apart from each other. We mark on those images the WISPR tracked features that can be identified in the COR2 images. 

Combining the EUV and coronagraph data, we associated features 0, 1, and 2 with the typical shapes that we are found in filaments during their evolution. Features 3 and 4 instead appear to be related to the core of the CME, indicated by the red arrow in the COR1 images (panels (A)-(D)). In Fig. \ref{overview} panel (C), we also mark with red arrows the thread patterns observed in C3 that might be related to features 3 and 4 as observed in WISPR images (panel D from Fig. \ref{overview} and shown in more detail in Fig.\ref{high_cadence_mosaic}).  

\section{Discussion and conclusions}\label{discussion}
The internal structure of CMEs has been explored in the past, but the new generation of white-light instruments, such as WISPR, deliver images with a much higher sensitivity thanks to the unprecedented orbit of PSP. These images hold interesting results, not yet fully explored regarding the characteristics of the MFR evolving into interplanetary space. 

We studied the CME event on December 8, 2022, related to an eruptive filament. The eruption occurred close to the western limb of the Sun and was observed in white-light by multiple spacecraft located at different distances and vantage points. While the STA and SOHO spacecraft, close to 1~AU and separated by 14°, remained almost constant in their position, PSP was close to the Sun and changed in distance from 0.16~AU to 0.11~AU and in longitude by $\sim$24° during the CME observing time of 32 hours. Using the fast motion of PSP, we investigated the WISPR data using a 3D reconstruction method, as described in Sect.~ \ref{method} and Appendix \ref{appendix}. We tracked 10 well-observed small-scale density structures in WISPR images (Figs. ~\ref{high_cadence_mosaic}-~\ref{wl_mosaic} and ~\ref{BLOBS_mosaic}). The results enabled us to categorize the tracked features and to locate them in the context of the global CME appearance derived from GCS reconstruction.

Based on their morphology, propagation direction, and dynamics we found three different groups of features. Group 1 covers a C- or J-shaped appearance (features 0, 1, 2, 5), group 2 are density patterns characterized by aligned threads (features 3, 4), and group 3 covers blobs (features 6, 7, 8, 9). The C- and J-shaped features can be interpreted as twisted plasma-containing magnetic field structures, which are commonly observed in CMEs. Concave-upward features are frequently observed in EUV. For example, in our study these features were observed by SUVI and EUVI (see the horns indicated with white arrows in Figs. ~\ref{polar_euv_304}-~\ref{euv_cor1_counterpart}), and they may be indicators of the extent of the flux rope cavity \citep{Vourlidas2013}. The EUV horns have an appearance similar to the C- and J-shapes observed in WISPR, which suggests coronal temperatures for these features. The high-density blobs observed are usually related to plasmoid instabilities in the post-CME current sheet \citep[see, e.g.,][]{Ciaravella2008,Schanche16, DiMatteo_2019, Poirier2023AA_model, Webb2016, Lee2020ApJ}. We found numerous blobs propagating in the rear part of the CME under study in a time window of approximately 22 hours. The blobs were derived to move slower by about 100\,km/s with respect to the features tracked at the front of the CME. 

The thread-like features (group 2) appear to be localized on a global fish-bone structure with a certain angle to the "backbone" of that structure. As these features trace the fine structure of the magnetic field, they might refer to some sheared field with a confined density enhancement. They likely resemble cool plasma material as observed in EUV at chromospheric temperatures (see red arrow in Figs. ~\ref{polar_euv_304}-~\ref{euv_cor1_counterpart}), evolving into localized structures during the filament eruption. This can be compared to other prominence eruption studies. For example, \cite{Su2015ApJ} reported a series of vertical threads aligned along an elongated structure together with an associated U-shape and blob features likely related to signatures of magnetic reconnection. In general, we note that the high-resolution data from WISPR allow to relate in more detail CME small-scale structures to typical elements of a filament \citep[see e.g.,][]{Parenti2014}. This enhanced level of detail might shed more light on the early evolution of the internal structure of CMEs and contributes to a better understanding of why such a low number of CMEs covering prominence material is detected in situ \citep[see e.g.,][]{Niembro_2023}.

When relating the tracked magnetic structures to the global CME structure, we found interesting results. To reconstruct the CME flux rope structure in 3D, we used the GCS model (see Fig.~ \ref{3d_view}). We obtained that all features of group 1 are clearly located within the CME's eastern part. We also found group 2 to be clearly located in the western leg of the CME. In contrast, the blobs from group 3 are distributed over a large longitudinal range of $\sim$40° and do not show a clustered behavior. With respect to the derived GCS geometry, all blobs (features 6--9) are located at the edges of the CME's eastern and western flanks, respectively. Features 8 and 9 are even outside of the reconstructed flux rope. This may hint toward either the spatial extent of the post-CME current sheet, and hence its three-dimensionality, or to the possible influence of the ambient corona in the evolutionary behavior of the CS \citep[e.g.,][]{Bemporad2006,Lin2015} eventually leading to a secondary source for the formation of the blobs. A clear distinction between these two processes is not possible due to the uncertainties related to both the PSP reconstruction and the GCS method (see Sect.~\ref{method}). Nevertheless, we encourage further studies that will relate small-scale density structures to the global CME appearance in order to better constrain their physical generation processes.

The combined views from SOHO, STA, and PSP allowed us to compare the 3D shape of the global CME appearance from 1~AU and from 0.11-0.16~AU. Interestingly, we found that WISPR does not detect the outer CME envelope, in contrast to the HI-1 images that clearly show the typical bubble-like front (see Fig.~\ref{gcs_hi}). Because PSP is much closer to the CME than SOHO and STA, the WISPR line of sight through the CME is much shorter than for the 1 AU imagers. Therefore, the broad outer front is "resolved out" in WISPR images. On the other hand, due to its closer proximity, WISPR gives higher signals for small-scale structures in the CME body \citep[see also e.g.,][]{Vourlidas2016,Nistico2019,Hess2020,Hess2021}. We also note that similar effects, namely a different appearance of coronal EUV waves at the solar surface as viewed from different vantage points, were explained by the integration of intensity along different lines of sight \citep[e.g.,][]{Patsourakos_Vourlidas_2009}.

We conclude that CME flux ropes might consist of different magnetic fine structures that evolve differently. These results are relevant for future CME propagation and forecasting models, especially those that implement magnetized flux ropes \citep[e.g.,][]{Torok2018,Verbeke2019}. To accurately interpret the observed density structures of a CME, which trace magnetic structures, it is essential to gather image data from various distances. This approach offers complementary perspectives on the CME's appearance, accounting for projection and line-of-sight integrated density effects that influence its visual presentation.

\bibliographystyle{aa} 

\begin{thebibliography}{84}
\expandafter\ifx\csname natexlab\endcsname\relax\def\natexlab#1{#1}\fi

\bibitem[{{Amari} {et~al.}(2003){Amari}, {Luciani}, {Aly}, {Mikic}, \& {Linker}}]{Amari2003ApJ_filament_fluxrope}
{Amari}, T., {Luciani}, J.~F., {Aly}, J.~J., {Mikic}, Z., \& {Linker}, J. 2003, \apj, 595, 1231

\bibitem[{{Astropy Collaboration} {et~al.}(2018){Astropy Collaboration}, {Price-Whelan}, {Sip{\H{o}}cz}, {G{\"u}nther}, {Lim}, {Crawford}, {Conseil}, {Shupe}, {Craig}, {Dencheva}, {Ginsburg}, {VanderPlas}, {Bradley}, {P{\'e}rez-Su{\'a}rez}, {de Val-Borro}, {Aldcroft}, {Cruz}, {Robitaille}, {Tollerud}, {Ardelean}, {Babej}, {Bach}, {Bachetti}, {Bakanov}, {Bamford}, {Barentsen}, {Barmby}, {Baumbach}, {Berry}, {Biscani}, {Boquien}, {Bostroem}, {Bouma}, {Brammer}, {Bray}, {Breytenbach}, {Buddelmeijer}, {Burke}, {Calderone}, {Cano Rodr{\'\i}guez}, {Cara}, {Cardoso}, {Cheedella}, {Copin}, {Corrales}, {Crichton}, {D'Avella}, {Deil}, {Depagne}, {Dietrich}, {Donath}, {Droettboom}, {Earl}, {Erben}, {Fabbro}, {Ferreira}, {Finethy}, {Fox}, {Garrison}, {Gibbons}, {Goldstein}, {Gommers}, {Greco}, {Greenfield}, {Groener}, {Grollier}, {Hagen}, {Hirst}, {Homeier}, {Horton}, {Hosseinzadeh}, {Hu}, {Hunkeler}, {Ivezi{\'c}}, {Jain}, {Jenness}, {Kanarek}, {Kendrew}, {Kern}, {Kerzendorf}, {Khvalko}, {King}, {Kirkby}, {Kulkarni},
  {Kumar}, {Lee}, {Lenz}, {Littlefair}, {Ma}, {Macleod}, {Mastropietro}, {McCully}, {Montagnac}, {Morris}, {Mueller}, {Mumford}, {Muna}, {Murphy}, {Nelson}, {Nguyen}, {Ninan}, {N{\"o}the}, {Ogaz}, {Oh}, {Parejko}, {Parley}, {Pascual}, {Patil}, {Patil}, {Plunkett}, {Prochaska}, {Rastogi}, {Reddy Janga}, {Sabater}, {Sakurikar}, {Seifert}, {Sherbert}, {Sherwood-Taylor}, {Shih}, {Sick}, {Silbiger}, {Singanamalla}, {Singer}, {Sladen}, {Sooley}, {Sornarajah}, {Streicher}, {Teuben}, {Thomas}, {Tremblay}, {Turner}, {Terr{\'o}n}, {van Kerkwijk}, {de la Vega}, {Watkins}, {Weaver}, {Whitmore}, {Woillez}, {Zabalza}, \& {Astropy Contributors}}]{astropy2018AJ}
{Astropy Collaboration}, {Price-Whelan}, A.~M., {Sip{\H{o}}cz}, B.~M., {et~al.} 2018, \aj, 156, 123

\bibitem[{{Aulanier} \& {Schmieder}(2002)}]{Aulanier2002}
{Aulanier}, G. \& {Schmieder}, B. 2002, \aap, 386, 1106

\bibitem[{{Bemporad} {et~al.}(2006){Bemporad}, {Poletto}, {Suess}, {Ko}, {Schwadron}, {Elliott}, \& {Raymond}}]{Bemporad2006}
{Bemporad}, A., {Poletto}, G., {Suess}, S.~T., {et~al.} 2006, \apj, 638, 1110

\bibitem[{{Bohlin} {et~al.}(1980){Bohlin}, {Frost}, {Burr}, {Guha}, \& {Withbroe}}]{SMM1980}
{Bohlin}, J.~D., {Frost}, K.~J., {Burr}, P.~T., {Guha}, A.~K., \& {Withbroe}, G.~L. 1980, \solphys, 65, 5

\bibitem[{{Braga} \& {Vourlidas}(2021)}]{Braga2021}
{Braga}, C.~R. \& {Vourlidas}, A. 2021, \aap, 650, A31

\bibitem[{{Braga} {et~al.}(2022){Braga}, {Vourlidas}, {Liewer}, {Hess}, {Stenborg}, \& {Riley}}]{Braga2022}
{Braga}, C.~R., {Vourlidas}, A., {Liewer}, P.~C., {et~al.} 2022, \apj, 938, 13

\bibitem[{{Brueckner} {et~al.}(1995){Brueckner}, {Howard}, {Koomen}, {Korendyke}, {Michels}, {Moses}, {Socker}, {Dere}, {Lamy}, {Llebaria}, {Bout}, {Schwenn}, {Simnett}, {Bedford}, \& {Eyles}}]{Brueckner_1995_lascoc2}
{Brueckner}, G.~E., {Howard}, R.~A., {Koomen}, M.~J., {et~al.} 1995, \solphys, 162, 357

\bibitem[{{Byrne} {et~al.}(2010){Byrne}, {Maloney}, {McAteer}, {Refojo}, \& {Gallagher}}]{Byrne2010}
{Byrne}, J.~P., {Maloney}, S.~A., {McAteer}, R.~T.~J., {Refojo}, J.~M., \& {Gallagher}, P.~T. 2010, Nature Communications, 1, 74

\bibitem[{{Chen}(2011)}]{Chen2011}
{Chen}, P.~F. 2011, Living Reviews in Solar Physics, 8, 1

\bibitem[{{Chen} {et~al.}(2014){Chen}, {Harra}, \& {Fang}}]{Chen2014ApJ}
{Chen}, P.~F., {Harra}, L.~K., \& {Fang}, C. 2014, \apj, 784, 50

\bibitem[{Ciaravella \& Raymond(2008)}]{Ciaravella2008}
Ciaravella, A. \& Raymond, J.~C. 2008, The Astrophysical Journal, 686, 1372

\bibitem[{Darnel {et~al.}(2022)Darnel, Seaton, Bethge, Rachmeler, Jarvis, Hill, Peck, Hughes, Shapiro, Riley, Vasudevan, Shing, Koener, Edwards, Mathur, \& Timothy}]{SUVI_GOES_DARNEL}
Darnel, J.~M., Seaton, D.~B., Bethge, C., {et~al.} 2022, Space Weather, 20, e2022SW003044, e2022SW003044 2022SW003044

\bibitem[{{DeVore} \& {Antiochos}(2000)}]{DeVore2000ApJ_filament_sheared_arcades}
{DeVore}, C.~R. \& {Antiochos}, S.~K. 2000, \apj, 539, 954

\bibitem[{{Di Matteo} {et~al.}(2019){Di Matteo}, {Viall}, {Kepko}, {Wallace}, {Arge}, \& {MacNeice}}]{DiMatteo_2019}
{Di Matteo}, S., {Viall}, N.~M., {Kepko}, L., {et~al.} 2019, Journal of Geophysical Research (Space Physics), 124, 837

\bibitem[{{Domingo} {et~al.}(1995){Domingo}, {Fleck}, \& {Poland}}]{Domingo_1995_soho}
{Domingo}, V., {Fleck}, B., \& {Poland}, A.~I. 1995, \solphys, 162, 1

\bibitem[{{Forbes}(2000)}]{Forbes2000}
{Forbes}, T.~G. 2000, \jgr, 105, 23153

\bibitem[{{Fox} {et~al.}(2016){Fox}, {Velli}, {Bale}, {Decker}, {Driesman}, {Howard}, {Kasper}, {Kinnison}, {Kusterer}, {Lario}, {Lockwood}, {McComas}, {Raouafi}, \& {Szabo}}]{Fox2016}
{Fox}, N.~J., {Velli}, M.~C., {Bale}, S.~D., {et~al.} 2016, \ssr, 204, 7

\bibitem[{{Gopalswamy}(2022)}]{Gopalswamy2022}
{Gopalswamy}, N. 2022, Atmosphere, 13, 1781

\bibitem[{{Green} {et~al.}(2018){Green}, {T{\"o}r{\"o}k}, {Vr{\v{s}}nak}, {Manchester}, \& {Veronig}}]{Green2018}
{Green}, L.~M., {T{\"o}r{\"o}k}, T., {Vr{\v{s}}nak}, B., {Manchester}, W., \& {Veronig}, A. 2018, \ssr, 214, 46

\bibitem[{{Guo} {et~al.}(2021){Guo}, {Zhou}, {Guo}, {Ni}, {Karpen}, \& {Chen}}]{Guo2021ApJ}
{Guo}, J.~H., {Zhou}, Y.~H., {Guo}, Y., {et~al.} 2021, \apj, 920, 131

\bibitem[{{Hess} {et~al.}(2021){Hess}, {Howard}, {Stenborg}, {Linton}, {Vourlidas}, {Thernisien}, {Colaninno}, {Rich}, {Wang}, {Battams}, \& {Kuroda}}]{Hess2021}
{Hess}, P., {Howard}, R.~A., {Stenborg}, G., {et~al.} 2021, \solphys, 296, 94

\bibitem[{{Hess} {et~al.}(2020){Hess}, {Rouillard}, {Kouloumvakos}, {Liewer}, {Zhang}, {Dhakal}, {Stenborg}, {Colaninno}, \& {Howard}}]{Hess2020}
{Hess}, P., {Rouillard}, A.~P., {Kouloumvakos}, A., {et~al.} 2020, \apjs, 246, 25

\bibitem[{{Howard} {et~al.}(2008){Howard}, {Moses}, {Vourlidas}, {Newmark}, {Socker}, {Plunkett}, {Korendyke}, {Cook}, {Hurley}, {Davila}, {Thompson}, {St Cyr}, {Mentzell}, {Mehalick}, {Lemen}, {Wuelser}, {Duncan}, {Tarbell}, {Wolfson}, {Moore}, {Harrison}, {Waltham}, {Lang}, {Davis}, {Eyles}, {Mapson-Menard}, {Simnett}, {Halain}, {Defise}, {Mazy}, {Rochus}, {Mercier}, {Ravet}, {Delmotte}, {Auchere}, {Delaboudiniere}, {Bothmer}, {Deutsch}, {Wang}, {Rich}, {Cooper}, {Stephens}, {Maahs}, {Baugh}, {McMullin}, \& {Carter}}]{Howard_2008_secchi}
{Howard}, R.~A., {Moses}, J.~D., {Vourlidas}, A., {et~al.} 2008, \ssr, 136, 67

\bibitem[{{Howard} {et~al.}(2022){Howard}, {Stenborg}, {Vourlidas}, {Gallagher}, {Linton}, {Hess}, {Rich}, \& {Liewer}}]{Howard2022}
{Howard}, R.~A., {Stenborg}, G., {Vourlidas}, A., {et~al.} 2022, \apj, 936, 43

\bibitem[{Howard {et~al.}(2023)Howard, Vourlidas, \& Stenborg}]{Howard_2023}
Howard, R.~A., Vourlidas, A., \& Stenborg, G. 2023, Frontiers in Astronomy and Space Sciences, 10

\bibitem[{{Hunter}(2007)}]{matplotlib2007}
{Hunter}, J.~D. 2007, Computing in Science and Engineering, 9, 90

\bibitem[{{Illing} \& {Hundhausen}(1983)}]{Illing_1983}
{Illing}, R.~M.~E. \& {Hundhausen}, A.~J. 1983, \jgr, 88, 10210

\bibitem[{{Kahler} \& {Webb}(2007)}]{kahler07}
{Kahler}, S.~W. \& {Webb}, D.~F. 2007, Journal of Geophysical Research (Space Physics), 112, A09103

\bibitem[{{Kaiser} {et~al.}(2008){Kaiser}, {Kucera}, {Davila}, {St. Cyr}, {Guhathakurta}, \& {Christian}}]{Kaiser_STEREO}
{Kaiser}, M.~L., {Kucera}, T.~A., {Davila}, J.~M., {et~al.} 2008, Space Science Review, 136, 5

\bibitem[{{Kouloumvakos} {et~al.}(2022){Kouloumvakos}, {Rodr{\'\i}guez-Garc{\'\i}a}, {Gieseler}, {Price}, {Vourlidas}, \& {Vainio}}]{Kouloumvakos2022}
{Kouloumvakos}, A., {Rodr{\'\i}guez-Garc{\'\i}a}, L., {Gieseler}, J., {et~al.} 2022, Frontiers in Astronomy and Space Sciences, 9, 974137

\bibitem[{Krimchansky {et~al.}(2004)Krimchansky, Machi, Cauffman, \& Davis}]{Krimchansky_goes2004}
Krimchansky, A., Machi, D., Cauffman, S., \& Davis, M. 2004, Proceedings of SPIE - The International Society for Optical Engineering, 5570

\bibitem[{{Lee} {et~al.}(2020){Lee}, {Cho}, {Lee}, {Cho}, {Lee}, {Miyashita}, {Kim}, {Kim}, \& {Jang}}]{Lee2020ApJ}
{Lee}, J.-O., {Cho}, K.-S., {Lee}, K.-S., {et~al.} 2020, \apj, 892, 129

\bibitem[{{Liewer} {et~al.}(2019){Liewer}, {Vourlidas}, {Thernisien}, {Qiu}, {Penteado}, {Nistic{\`o}}, {Howard}, \& {Bothmer}}]{Liewer2019}
{Liewer}, P., {Vourlidas}, A., {Thernisien}, A., {et~al.} 2019, \solphys, 294, 93

\bibitem[{{Liewer} {et~al.}(2022){Liewer}, {Qiu}, {Ark}, {Penteado}, {Stenborg}, {Vourlidas}, {Hall}, \& {Riley}}]{Liewer2022}
{Liewer}, P.~C., {Qiu}, J., {Ark}, F., {et~al.} 2022, \solphys, 297, 128

\bibitem[{{Liewer} {et~al.}(2020){Liewer}, {Qiu}, {Penteado}, {Hall}, {Vourlidas}, \& {Howard}}]{Liewer2020}
{Liewer}, P.~C., {Qiu}, J., {Penteado}, P., {et~al.} 2020, \solphys, 295, 140

\bibitem[{{Liewer} {et~al.}(2021){Liewer}, {Qiu}, {Vourlidas}, {Hall}, \& {Penteado}}]{Liewer2021}
{Liewer}, P.~C., {Qiu}, J., {Vourlidas}, A., {Hall}, J.~R., \& {Penteado}, P. 2021, \aap, 650, A32

\bibitem[{{Lin} {et~al.}(2015){Lin}, {Murphy}, {Shen}, {Raymond}, {Reeves}, {Zhong}, {Wu}, \& {Li}}]{Lin2015}
{Lin}, J., {Murphy}, N.~A., {Shen}, C., {et~al.} 2015, \ssr, 194, 237

\bibitem[{{Lugaz} {et~al.}(2009){Lugaz}, {Vourlidas}, \& {Roussev}}]{lugaz09}
{Lugaz}, N., {Vourlidas}, A., \& {Roussev}, I.~I. 2009, Annales Geophysicae, 27, 3479

\bibitem[{{Mierla} {et~al.}(2010){Mierla}, {Inhester}, {Antunes}, {Boursier}, {Byrne}, {Colaninno}, {Davila}, {de Koning}, {Gallagher}, {Gissot}, {Howard}, {Howard}, {Kramar}, {Lamy}, {Liewer}, {Maloney}, {Marqu{\'e}}, {McAteer}, {Moran}, {Rodriguez}, {Srivastava}, {St. Cyr}, {Stenborg}, {Temmer}, {Thernisien}, {Vourlidas}, {West}, {Wood}, \& {Zhukov}}]{Mierla2010}
{Mierla}, M., {Inhester}, B., {Antunes}, A., {et~al.} 2010, Annales Geophysicae, 28, 203

\bibitem[{{Mishra} \& {Teriaca}(2023)}]{Mishra2023}
{Mishra}, W. \& {Teriaca}, L. 2023, Journal of Astrophysics and Astronomy, 44, 20

\bibitem[{{Moore} {et~al.}(2001){Moore}, {Sterling}, {Hudson}, \& {Lemen}}]{Moore2001}
{Moore}, R.~L., {Sterling}, A.~C., {Hudson}, H.~S., \& {Lemen}, J.~R. 2001, \apj, 552, 833

\bibitem[{{M{\"u}ller} {et~al.}(2017){M{\"u}ller}, {Nicula}, {Felix}, {Verstringe}, {Bourgoignie}, {Csillaghy}, {Berghmans}, {Jiggens}, {Garc{\'\i}a-Ortiz}, {Ireland}, {Zahniy}, \& {Fleck}}]{Jhelioviewer_2017}
{M{\"u}ller}, D., {Nicula}, B., {Felix}, S., {et~al.} 2017, \aap, 606, A10

\bibitem[{{Niembro} {et~al.}(2023){Niembro}, {Seaton}, {Hess}, {Berghmans}, {Andretta}, {Reeves}, {Riley}, {Stevens}, {Landini}, {Sasso}, {Verbeeck}, {Susino}, \& {Uslenghi}}]{Niembro_2023}
{Niembro}, T., {Seaton}, D.~B., {Hess}, P., {et~al.} 2023, Frontiers in Astronomy and Space Sciences, 10, 1191294

\bibitem[{{Nistic{\`o}} {et~al.}(2019){Nistic{\`o}}, {Bothmer}, {Liewer}, {Vourlidas}, \& {Thernisien}}]{Nistico2019}
{Nistic{\`o}}, G., {Bothmer}, V., {Liewer}, P., {Vourlidas}, A., \& {Thernisien}, A. 2019, Nuovo Cimento C Geophysics Space Physics C, 42, 21

\bibitem[{{Okamoto} {et~al.}(2007){Okamoto}, {Tsuneta}, {Berger}, {Ichimoto}, {Katsukawa}, {Lites}, {Nagata}, {Shibata}, {Shimizu}, {Shine}, {Suematsu}, {Tarbell}, \& {Title}}]{Okamoto2007}
{Okamoto}, T.~J., {Tsuneta}, S., {Berger}, T.~E., {et~al.} 2007, Science, 318, 1577

\bibitem[{{Parenti}(2014)}]{Parenti2014}
{Parenti}, S. 2014, Living Reviews in Solar Physics, 11, 1

\bibitem[{{Patel} {et~al.}(2023){Patel}, {West}, {Seaton}, {Hess}, {Niembro}, \& {Reeves}}]{Patel_2023ApJ}
{Patel}, R., {West}, M.~J., {Seaton}, D.~B., {et~al.} 2023, \apjl, 955, L1

\bibitem[{Patsourakos \& Vourlidas(2009)}]{Patsourakos_Vourlidas_2009}
Patsourakos, S. \& Vourlidas, A. 2009, Astrophysical Journal, 700, L182–L186

\bibitem[{{Poirier} {et~al.}(2023){Poirier}, {R{\'e}ville}, {Rouillard}, {Kouloumvakos}, \& {Valette}}]{Poirier2023AA_model}
{Poirier}, N., {R{\'e}ville}, V., {Rouillard}, A.~P., {Kouloumvakos}, A., \& {Valette}, E. 2023, \aap, 677, A108

\bibitem[{{Pulkkinen}(2007)}]{Pulkkinen2007}
{Pulkkinen}, T. 2007, Living Reviews in Solar Physics, 4, 1

\bibitem[{{Raouafi} {et~al.}(2023){Raouafi}, {Matteini}, {Squire}, {Badman}, {Velli}, {Klein}, {Chen}, {Matthaeus}, {Szabo}, {Linton}, {Allen}, {Szalay}, {Bruno}, {Decker}, {Akhavan-Tafti}, {Agapitov}, {Bale}, {Bandyopadhyay}, {Battams}, {Ber{\v{c}}i{\v{c}}}, {Bourouaine}, {Bowen}, {Cattell}, {Chandran}, {Chhiber}, {Cohen}, {D'Amicis}, {Giacalone}, {Hess}, {Howard}, {Horbury}, {Jagarlamudi}, {Joyce}, {Kasper}, {Kinnison}, {Laker}, {Liewer}, {Malaspina}, {Mann}, {McComas}, {Niembro-Hernandez}, {Nieves-Chinchilla}, {Panasenco}, {Pokorn{\'y}}, {Pusack}, {Pulupa}, {Perez}, {Riley}, {Rouillard}, {Shi}, {Stenborg}, {Tenerani}, {Verniero}, {Viall}, {Vourlidas}, {Wood}, {Woodham}, \& {Woolley}}]{Raouafi2023}
{Raouafi}, N.~E., {Matteini}, L., {Squire}, J., {et~al.} 2023, \ssr, 219, 8

\bibitem[{{Reback} {et~al.}(2020){Reback}, {McKinney}, {Jbrockmendel}, {Van Den Bossche}, {Augspurger}, {Cloud}, {Gfyoung}, {Sinhrks}, {Klein}, {Hawkins}, {Roeschke}, {Tratner}, {She}, {Ayd}, {Petersen}, {MomIsBestFriend}, {Garcia}, {Schendel}, {Hayden}, {Jancauskas}, {Battiston}, {Saxton}, {Seabold}, {Alimcmaster1}, {Chris-B1}, {H-Vetinari}, {Hoyer}, {Dong}, {Overmeire}, \& {Winkel}}]{pandas2020}
{Reback}, J., {McKinney}, W., {Jbrockmendel}, {et~al.} 2020, {pandas-dev/pandas: Pandas 1.0.5}

\bibitem[{{Riley} {et~al.}(2007){Riley}, {Lionello}, {Miki{\'c}}, {Linker}, {Clark}, {Lin}, \& {Ko}}]{Riley_2007_bursty_reconnection}
{Riley}, P., {Lionello}, R., {Miki{\'c}}, Z., {et~al.} 2007, \apj, 655, 591

\bibitem[{{Rollett} {et~al.}(2016){Rollett}, {M{\"o}stl}, {Isavnin}, {Davies}, {Kubicka}, {Amerstorfer}, \& {Harrison}}]{rollett2016ApJ}
{Rollett}, T., {M{\"o}stl}, C., {Isavnin}, A., {et~al.} 2016, \apj, 824, 131

\bibitem[{{Romeo} {et~al.}(2023){Romeo}, {Braga}, {Badman}, {Larson}, {Stevens}, {Huang}, {Phan}, {Rahmati}, {Livi}, {Alnussirat}, {Whittlesey}, {Szabo}, {Klein}, {Niembro-Hernandez}, {Paulson}, {Verniero}, {Lario}, {Raouafi}, {Ervin}, {Kasper}, {Pulupa}, {Bale}, \& {Linton}}]{Romeo2023ApJ}
{Romeo}, O.~M., {Braga}, C.~R., {Badman}, S.~T., {et~al.} 2023, \apj, 954, 168

\bibitem[{{Ruan} {et~al.}(2018){Ruan}, {Schmieder}, {Mein}, {Mein}, {Labrosse}, {Gun{\'a}r}, \& {Chen}}]{Ruan2018}
{Ruan}, G., {Schmieder}, B., {Mein}, P., {et~al.} 2018, \apj, 865, 123

\bibitem[{{Schanche} {et~al.}(2016){Schanche}, {Reeves}, \& {Webb}}]{Schanche16}
{Schanche}, N.~E., {Reeves}, K.~K., \& {Webb}, D.~F. 2016, \apj, 831, 47

\bibitem[{{Schmieder} {et~al.}(2013){Schmieder}, {D{\'e}moulin}, \& {Aulanier}}]{Schmieder2013}
{Schmieder}, B., {D{\'e}moulin}, P., \& {Aulanier}, G. 2013, Advances in Space Research, 51, 1967

\bibitem[{{Schmieder} {et~al.}(2014){Schmieder}, {Tian}, {Kucera}, {L{\'o}pez Ariste}, {Mein}, {Mein}, {Dalmasse}, \& {Golub}}]{Schmieder2014}
{Schmieder}, B., {Tian}, H., {Kucera}, T., {et~al.} 2014, \aap, 569, A85

\bibitem[{{Su} {et~al.}(2015){Su}, {van Ballegooijen}, {McCauley}, {Ji}, {Reeves}, \& {DeLuca}}]{Su2015ApJ}
{Su}, Y., {van Ballegooijen}, A., {McCauley}, P., {et~al.} 2015, \apj, 807, 144

\bibitem[{{SunPy Community} {et~al.}(2020){SunPy Community}, {Barnes}, {Bobra}, {Christe}, {Freij}, {Hayes}, {Ireland}, {Mumford}, {Perez-Suarez}, {Ryan}, {Shih}, {Chanda}, {Glogowski}, {Hewett}, {Hughitt}, {Hill}, {Hiware}, {Inglis}, {Kirk}, {Konge}, {Mason}, {Maloney}, {Murray}, {Panda}, {Park}, {Pereira}, {Reardon}, {Savage}, {Sip{\H{o}}cz}, {Stansby}, {Jain}, {Taylor}, {Yadav}, {Rajul}, \& {Dang}}]{sunpy2020ApJ}
{SunPy Community}, {Barnes}, W.~T., {Bobra}, M.~G., {et~al.} 2020, \apj, 890, 68

\bibitem[{{Tandberg-Hanssen}(1995)}]{Tandberg-Hanssen1995}
{Tandberg-Hanssen}, E. 1995, {The nature of solar prominences}, Vol. 199

\bibitem[{{Temmer}(2021)}]{Temmer2021}
{Temmer}, M. 2021, Living Reviews in Solar Physics, 18, 4

\bibitem[{{Temmer} \& {Bothmer}(2022)}]{TB22}
{Temmer}, M. \& {Bothmer}, V. 2022, \aap, 665, A70

\bibitem[{{Temmer} {et~al.}(2009){Temmer}, {Preiss}, \& {Veronig}}]{Temmer2009}
{Temmer}, M., {Preiss}, S., \& {Veronig}, A.~M. 2009, \solphys, 256, 183

\bibitem[{{Temmer} {et~al.}(2023){Temmer}, {Scolini}, {Richardson}, {Heinemann}, {Paouris}, {Vourlidas}, {Bisi}, {writing teams}, {:}, {Al-Haddad}, {Amerstorfer}, {Barnard}, {Buresova}, {Hofmeister}, {Iwai}, {Jackson}, {Jarolim}, {Jian}, {Linker}, {Lugaz}, {Manoharan}, {Mays}, {Mishra}, {Owens}, {Palmerio}, {Perri}, {Pomoell}, {Pinto}, {Samara}, {Singh}, {Sur}, {Verbeke}, {Veronig}, \& {Zhuang}}]{Temmer2023}
{Temmer}, M., {Scolini}, C., {Richardson}, I.~G., {et~al.} 2023, arXiv e-prints, arXiv:2308.04851

\bibitem[{{Thernisien}(2011)}]{Thernisien2011}
{Thernisien}, A. 2011, \apjs, 194, 33

\bibitem[{{Thernisien} {et~al.}(2009){Thernisien}, {Vourlidas}, \& {Howard}}]{Thernisien2009}
{Thernisien}, A., {Vourlidas}, A., \& {Howard}, R.~A. 2009, \solphys, 256, 111

\bibitem[{{T{\"o}r{\"o}k} {et~al.}(2018){T{\"o}r{\"o}k}, {Downs}, {Linker}, {Lionello}, {Titov}, {Miki{\'c}}, {Riley}, {Caplan}, \& {Wijaya}}]{Torok2018}
{T{\"o}r{\"o}k}, T., {Downs}, C., {Linker}, J.~A., {et~al.} 2018, \apj, 856, 75

\bibitem[{{van der Walt} {et~al.}(2011){van der Walt}, {Colbert}, \& {Varoquaux}}]{numpy2011}
{van der Walt}, S., {Colbert}, S.~C., \& {Varoquaux}, G. 2011, Computing in Science and Engineering, 13, 22

\bibitem[{{Verbeke} {et~al.}(2023){Verbeke}, {Mays}, {Kay}, {Riley}, {Palmerio}, {Dumbovi{\'c}}, {Mierla}, {Scolini}, {Temmer}, {Paouris}, {Balmaceda}, {Cremades}, \& {Hinterreiter}}]{Verbeke2023}
{Verbeke}, C., {Mays}, M.~L., {Kay}, C., {et~al.} 2023, Advances in Space Research, 72, 5243

\bibitem[{{Verbeke} {et~al.}(2019){Verbeke}, {Pomoell}, \& {Poedts}}]{Verbeke2019}
{Verbeke}, C., {Pomoell}, J., \& {Poedts}, S. 2019, \aap, 627, A111

\bibitem[{{Virtanen} {et~al.}(2020){Virtanen}, {Gommers}, {Oliphant}, {Haberland}, {Reddy}, {Cournapeau}, {Burovski}, {Peterson}, {Weckesser}, {Bright}, {van der Walt}, {Brett}, {Wilson}, {Millman}, {Mayorov}, {Nelson}, {Jones}, {Kern}, {Larson}, {Carey}, {Polat}, {Feng}, {Moore}, {VanderPlas}, {Laxalde}, {Perktold}, {Cimrman}, {Henriksen}, {Quintero}, {Harris}, {Archibald}, {Ribeiro}, {Pedregosa}, {van Mulbregt}, \& {SciPy 1. 0 Contributors}}]{scipy2020}
{Virtanen}, P., {Gommers}, R., {Oliphant}, T.~E., {et~al.} 2020, Nature Methods, 17, 261

\bibitem[{{Von Forstner}(2021)}]{Forstner21}
{Von Forstner}, J. L.~F. 2021, {johan12345/gcs\_python: Release 0.2.2}

\bibitem[{{Vourlidas} {et~al.}(2016){Vourlidas}, {Howard}, {Plunkett}, {Korendyke}, {Thernisien}, {Wang}, {Rich}, {Carter}, {Chua}, {Socker}, {Linton}, {Morrill}, {Lynch}, {Thurn}, {Van Duyne}, {Hagood}, {Clifford}, {Grey}, {Velli}, {Liewer}, {Hall}, {DeJong}, {Mikic}, {Rochus}, {Mazy}, {Bothmer}, \& {Rodmann}}]{Vourlidas2016}
{Vourlidas}, A., {Howard}, R.~A., {Plunkett}, S.~P., {et~al.} 2016, \ssr, 204, 83

\bibitem[{{Vourlidas} {et~al.}(2013){Vourlidas}, {Lynch}, {Howard}, \& {Li}}]{Vourlidas2013}
{Vourlidas}, A., {Lynch}, B.~J., {Howard}, R.~A., \& {Li}, Y. 2013, \solphys, 284, 179

\bibitem[{{Vr{\v{s}}nak}(2001)}]{Vrsnak2001}
{Vr{\v{s}}nak}, B. 2001, \solphys, 202, 173

\bibitem[{{Webb} {et~al.}(2003){Webb}, {Burkepile}, {Forbes}, \& {Riley}}]{Webb2003}
{Webb}, D.~F., {Burkepile}, J., {Forbes}, T.~G., \& {Riley}, P. 2003, Journal of Geophysical Research (Space Physics), 108, 1440

\bibitem[{{Webb} \& {Cliver}(1995)}]{Webb1995}
{Webb}, D.~F. \& {Cliver}, E.~W. 1995, \jgr, 100, 5853

\bibitem[{{Webb} \& {Vourlidas}(2016)}]{Webb2016}
{Webb}, D.~F. \& {Vourlidas}, A. 2016, \solphys, 291, 3725

\bibitem[{{Wood} {et~al.}(2009){Wood}, {Howard}, {Plunkett}, \& {Socker}}]{wood09}
{Wood}, B.~E., {Howard}, R.~A., {Plunkett}, S.~P., \& {Socker}, D.~G. 2009, \apj, 694, 707

\bibitem[{{Wuelser} {et~al.}(2004){Wuelser}, {Lemen}, {Tarbell}, {Wolfson}, {Cannon}, {Carpenter}, {Duncan}, {Gradwohl}, {Meyer}, {Moore}, {Navarro}, {Pearson}, {Rossi}, {Springer}, {Howard}, {Moses}, {Newmark}, {Delaboudiniere}, {Artzner}, {Auchere}, {Bougnet}, {Bouyries}, {Bridou}, {Clotaire}, {Colas}, {Delmotte}, {Jerome}, {Lamare}, {Mercier}, {Mullot}, {Ravet}, {Song}, {Bothmer}, \& {Deutsch}}]{Wuelser04}
{Wuelser}, J.-P., {Lemen}, J.~R., {Tarbell}, T.~D., {et~al.} 2004, in Society of Photo-Optical Instrumentation Engineers (SPIE) Conference Series, Vol. 5171, Telescopes and Instrumentation for Solar Astrophysics, ed. S.~{Fineschi} \& M.~A. {Gummin}, 111--122

\bibitem[{{Zhang} {et~al.}(2021){Zhang}, {Temmer}, {Gopalswamy}, {Malandraki}, {Nitta}, {Patsourakos}, {Shen}, {Vr{\v{s}}nak}, {Wang}, {Webb}, {Desai}, {Dissauer}, {Dresing}, {Dumbovi{\'c}}, {Feng}, {Heinemann}, {Laurenza}, {Lugaz}, \& {Zhuang}}]{Zhang2021}
{Zhang}, J., {Temmer}, M., {Gopalswamy}, N., {et~al.} 2021, Progress in Earth and Planetary Science, 8, 56

\end{thebibliography}

\newpage

\appendix
\section{Single-spacecraft reconstruction fits}\label{appendix}
As described in Sect.~\ref{method}, PSP on its own can be used to obtain the kinematic information and position of features belonging to an observed CME. The CME's small-scale structures we investigated are too faint to be revealed and tracked through the WISPR-I and WISPR-O FoVs. Therefore, we use the so-called LW-processed images \citep[described in detail in Appendix A in][]{Howard2022}. This approach exploits the time domain at each pixel location to obtain a smooth background (at the fifth-percentile level) in a time interval centered at the time of each image.

\begin{figure}[!hbt]
\centering
\includegraphics[scale = 0.7]{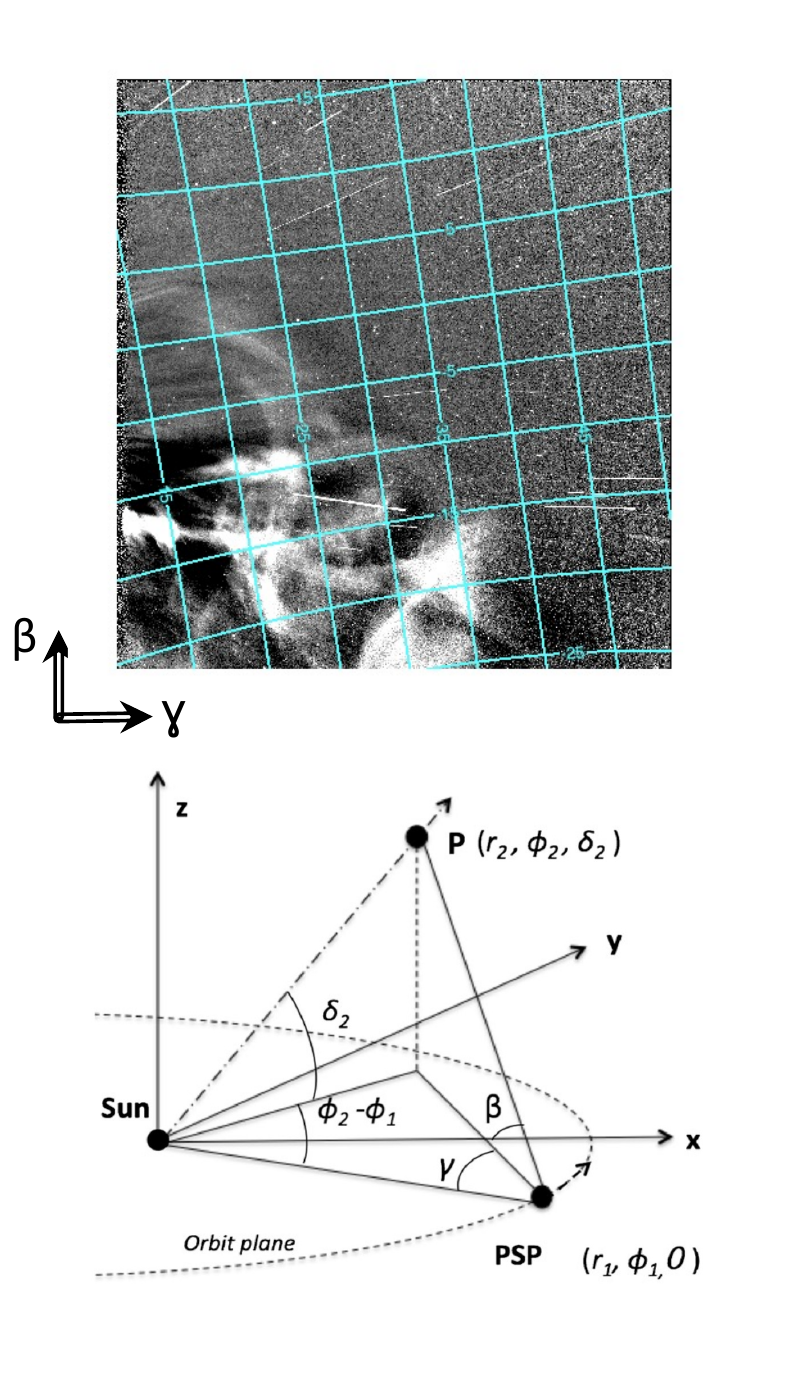}
\caption{Observational and orbit geometry parameters. Top: Example of the grid used by the IDL routine \texttt{wispr\_camera\_coords} to measure the projected angles in WISPR images. Respectively, the angle out of the plane is $\beta$ and the elongation is $\gamma$. Bottom: Schematic of the geometry used taken from \citep{Liewer2019} where PSP's orbit is approximated to be in the solar equatorial plane and through which we derived the geometrical equations (Eqs.~(~\ref{eq:geometic1})-~(\ref{eq:geometic2})).}
\label{methodIDL}
\end {figure}

 Figure~\ref{methodIDL} describes the geometry for reconstructing trajectories of small-scale structures in the WISPR FoV. We are interested in their location with respect to the Sun at a specific time t, given by $r_{2}(t)$. In addition, we wanted to obtain for each feature its longitude $\phi_{2}$ and its latitude $\delta_{2}$ in a heliocentric solar equatorial coordinate system, such as the HCI reference frame. Approximating PSP's orbit to be in the solar equatorial plane, its coordinates are given by [$r_{\rm PSP}, \phi_{\rm PSP}, \delta_{\rm PSP}]=[r_1, \phi_1, 0]$. PSP observes a point-like feature in the WISPR FoV that is identified by two projected angles: the elongation $\gamma$ in PSP's orbital plane and the angle out of this plane $\beta$, as illustrated in Fig.~\ref{methodIDL}. These angles were obtained by tracking the feature over time and converting its coordinate from pixel to PSP orbital frame coordinates using the SolarSoft/IDL routine \texttt{wispr\_camera\_coords} while accounting for effects of spacecraft location, projection, and distortion of the instruments \citep[see more details in][]{Braga2021}. Each of the identified features was tracked at least three times to ensure the track of the same feature was correctly followed over time. This allowed us to estimate the uncertainties in the results obtained according to the approach followed by \cite{Braga2021}. From these different tracks, \cite{Liewer2020} obtained the average $\beta$ and $\gamma$ at each time, and estimated the uncertainties in the $\beta$ and $\gamma$ measurements, and used these uncertainties to estimate the errors in the parameters of the reconstructed  trajectory of a feature.

From the measured $\gamma(t)$ and $\beta(t)$ of a specific feature, we determined the unknowns $r_2$, defined as $r_2(t)=r_0+v_0*t$ (with $v_0$ and $r_0$ as its constant speed and position at the initial time $t_0$); $\phi_2$; and $\delta_2$ by fitting the following two geometrical equations:
\begin{equation}
\beta (t) = {\rm atan} \bigg(\frac{\tan\delta_2\sin(\gamma(t))}{\sin[\phi_2-\phi_1(t)]}\bigg) ~ 
\label{eq:geometic1_simply}
\end{equation}

\begin{equation}
    \centering
    \cot(\gamma(t))=\frac{r_1(t)-r_2(t)\cos\delta_2\cos([\phi_2-\phi_1(t)])}{r_2(t)\cos\delta_2\sin([\phi_2-\phi_1(t)])} ~ .
    \label{eq:geometic2_simply}
\end{equation}

\noindent 
These two equations provide the conversion between the observed positions ($\beta$ and $\gamma$) of a feature with respect to WISPR and the feature's 3D position in the heliocentric inertia reference frame, assuming that PSP's orbit is in the solar equatorial plane. We note that PSP's orbit is inclined to the solar equator by a small angle, $\epsilon \approx $ 4°.  The above equations were therefore modified to include the correction due to the inclination $\epsilon$:

\begin{equation}
\beta (t) = {\rm atan} \bigg(\frac{\tan\delta_2\sin(\gamma(t))}{\sin[\phi_2-\phi_1(t)]}(1-F\sin\epsilon)\bigg) ~ 
\label{eq:geometic1}
\end{equation}

\begin{equation}
    \centering
    \cot(\gamma(t))=\frac{r_1(t)-r_2(t)\cos\delta_2\cos([\phi_2-\phi_1(t)])}{r_2(t)\cos\delta_2\sin([\phi_2-\phi_1(t)])}(1-G\sin\delta_2\sin\epsilon) ~ ,
    \label{eq:geometic2}
\end{equation}

\noindent where $G$ and $F$ specify the first order corrections, and their expressions are given in the appendix of \cite{Liewer2020}. With PSP's position ($\phi_1$, $r_1$, and $\epsilon$) known and assuming a radial motion at a constant speed, a feature's trajectory ($\phi_2, \delta_2, r_0$, and $v_0$) could be determined from a sequence of $\gamma(t)$ and $\beta(t)$ measurements.

So far, two different approaches have been adopted by the scientific community to solve these equations and obtain the unknowns. \cite{Liewer2020} solves the equations containing the $\epsilon$ term, Eqs.~(\ref{eq:geometic1})-~(\ref{eq:geometic2}),  performing a Levenberg–Marquardt least-squares fit \citep[see Appendix B of][for a discussion on deriving initial guess parameters]{Liewer2020}. \cite{Braga2021} solves the simplified equations and finds the best $\phi_2(t)$ and $\delta_2(t)$ by substituting simultaneously in Eqs.~(\ref{eq:geometic1_simply})-~(\ref{eq:geometic2_simply}) the initial guess for each unknown, which the user sets to vary within a certain range a priori. The results are then converted to the system of reference of the spacecraft in order to consider the inclination of PSP to the solar equatorial plane. Once the equations are solved per each set of initial guesses, the best fit is computed using the residuals. Indeed, the solution is the one that allows the $\beta_{der}$ and $\gamma_{der}$ to be derived through Eqs.~(\ref{eq:geometic1_simply})-~ (\ref{eq:geometic2_simply}), with the smallest deviations with respect to the measured $\beta_{mes}$ and $\gamma_{mes}$ values. In this study, we used both of the above-described approaches. 

\begin{figure*}
\centering
\includegraphics[scale = 0.60]{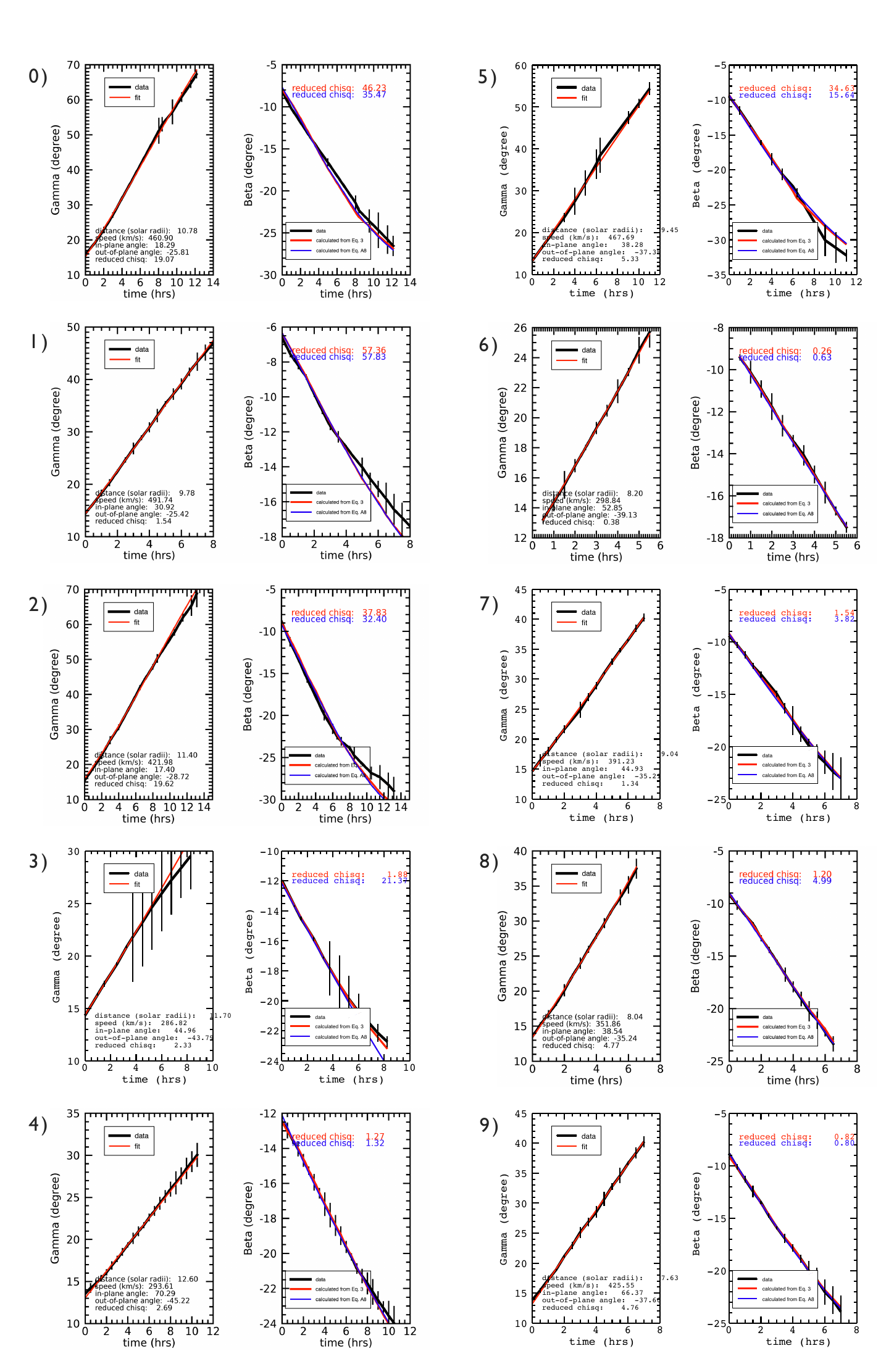}
\caption{Results obtained for features (0 to 9) using the approach of \citep{Liewer2020}. }\label{appendix_liewer}
\end{figure*}

\begin{figure*}
\centering
    \includegraphics[scale = 0.59]{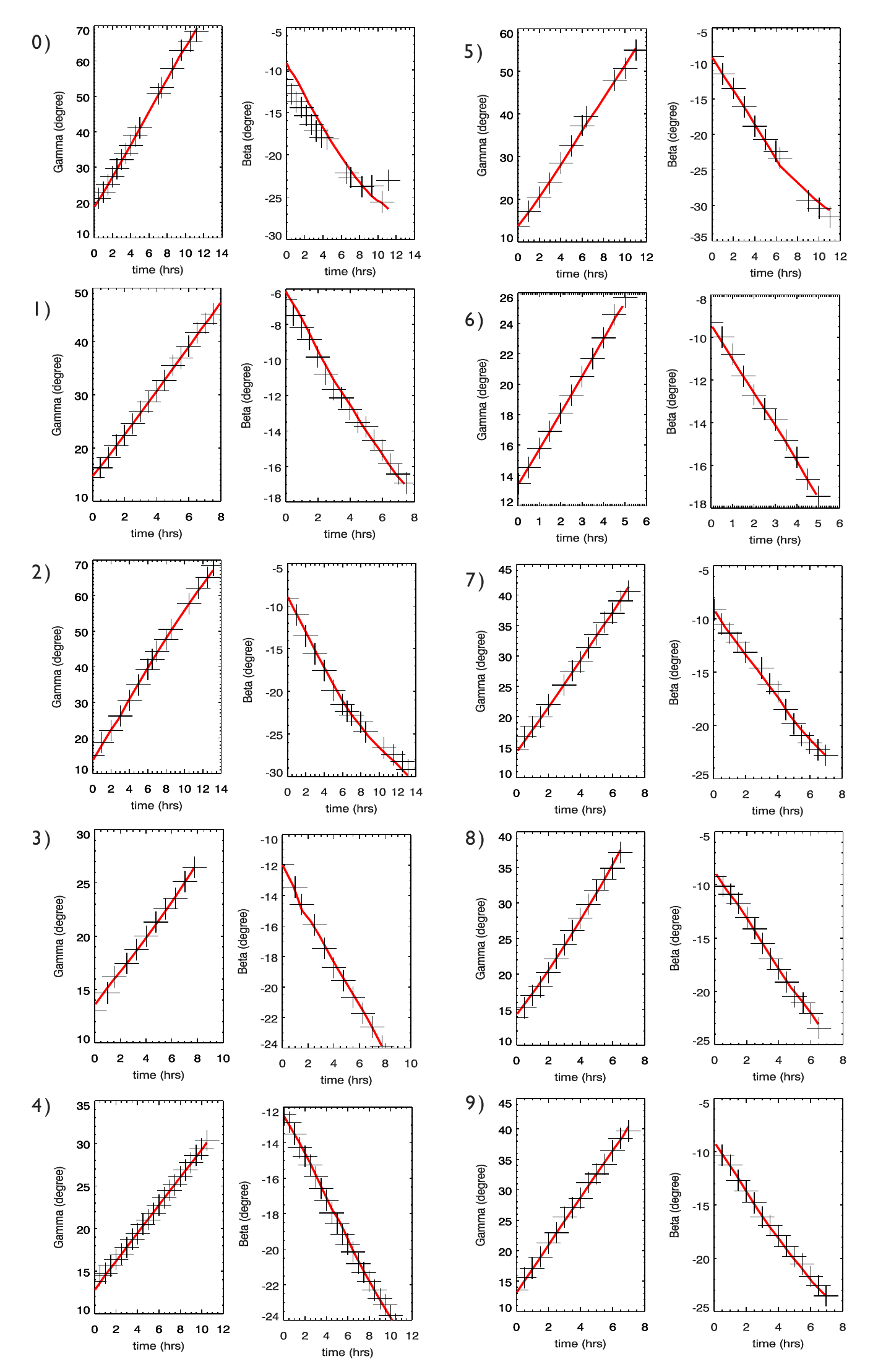}
\caption{Results obtained for features (0 to 9) using the approach of \citep{Braga2021}. The black crosses are the measured angles $\gamma$ and $\beta$, while the solid redline indicates the best fit of the equations discussed in Appendix \ref{appendix}, with the parameters shown in Table \ref{results_table}. }\label{appendix_braga}
\end{figure*}

Figures ~\ref{appendix_liewer} and \ref{appendix_braga} respectively show the calculation steps and plots related to solving Eqs.~(\ref{eq:geometic1})-~(\ref{eq:geometic2}) using the two different approaches by \cite{Liewer2020} and \cite{Braga2021} for all tracked features (0--9). The features were carefully selected considering a significant variation of the longitude of PSP during its observation in order to satisfy the condition for applying the reconstruction methods (Sect. ~\ref{method}). 

Figure \ref{appendix_liewer} shows for each feature the plots of the evolution of the elongation $\gamma$ and $\beta$ over time. The measured angles and their 1-$\sigma$ uncertainties, which were calculated considering the tracking performed three times per each substructure, are shown in black. In red we show the computed $\gamma$ and $\beta$ using the best-fitting parameters derived by forward fitting Eqs. (\ref{eq:geometic1})-~(\ref{eq:geometic2}). The $\beta$ angle can be also computed with a different equation \citep[see][]{Liewer2020}, which is shown in blue. 

Figure \ref{appendix_braga} shows per each feature the evolution of the elongation, $\gamma$, and angle out of the plane, $\beta$. The black crosses indicate the measurements performed on the angles, while the solid red line shows the best fit. The parameters for the fit were chosen by calculating the minimum of the residual function. We clarify that \cite{Braga2021} performed the fit on each of the tracks and that the best parameters, shown in Table~\ref{results_table}, are the average of the results obtained in the total number of tracks. Specifically, Fig.\ref{appendix_braga} only shows the evolution of the angles in one of the three tracks recorded for each feature. In constrast, \cite{Liewer2020} calculated the average for measurements and then performed the fit. This explains why in Fig.~\ref{appendix_liewer} the error bars of each measurement are shown.

In this study, the trajectories of the tracked features were determined with two approaches, and in both cases, we assumed that a feature moves radially with a constant speed, namely, $\phi_2$, $\delta_2$, $v_0$ are all constant. A relatively large deviation of the computed $\gamma$ and $\beta$ from the measurements, as shown in a few panels in \textcolor{red}{Fig.}~\ref{appendix_liewer}, may indicate that the motion of some of these features is not exactly radial at times. We will discuss these deviations in future work. The experiments on deriving time-dependent longitude and latitude are described by \citep{Braga2021}. 

\begin{acknowledgements}
G.M.C. and M.T. acknowledge support from the Young Researchers Program (YRP), project number AVO165300016. A.V and G.S are supported by WISPR Phase-E funds. C.R.B. acknowledges the support from the NASA STEREO/SECCHI (NNG17PP27I) program and NASA HGI grant 80NSSC23K0412. The work of P.C.L. was conducted at the Jet Propulsion Laboratory, California Institute of Technology under a contract from NASA. J. Q. is supported by NASA HGI grant 80NSSC22K0519. A.K. acknowledges financial support from NASA HTM grant 80NSSC24K0071. V.B. acknowledges funding of the CGAUSS project as German contribution to the WISPR camera on Parker Solar Probe by the German Space Agency through the German Bundesministerium für Wirtschaft und Klimaschutz under FKZ:  50OL2301.
G.M.C. thanks Iulia Ana Maria Chifu for fruitful discussions on the WISPR field of view, as well as Paulo Penteado for the support on the SolarSoft/WISPR IDL package.
Parker Solar Probe was designed, built, and is now operated by the Johns Hopkins Applied Physics Laboratory as part of NASA’s Living with a Star (LWS) program (contract NNN06AA01C). Support from the LWS management and technical team has played a critical role in the success of the Parker Solar Probe mission. The Wide-Field Imager for Parker Solar Probe (WISPR) instrument was designed, built, and is now operated by the US Naval Research Laboratory in collaboration with Johns Hopkins University/Applied Physics Laboratory, California Institute of Technology/Jet Propulsion Laboratory, University of Gottingen, Germany, Centre Spatiale de Liege, Belgium and University of Toulouse/Research Institute in Astrophysics and Planetology. The SECCHI data are produced by an international consortium of the NRL, LMSAL, and NASA GSFC (USA), RAL and U. Bham (UK), MPS (Germany), CSL (Belgium), IOTA, and IAS (France). The SOHO/LASCO data used here are produced by a consortium of the Naval Research Laboratory (USA), Max-Planck-Institut fuer Aeronomie (Germany), Laboratoire d’Astronomie (France), and the University of Birmingham (UK). SOHO is a project of international cooperation between ESA and NASA. SUVI was designed and built at Lockheed-Martin’s Advanced Technology Center in Palo Alto, California. 
\end{acknowledgements}
Software: JHelioviewer \footnote{\url{https://www.jhelioviewer.org/}}, has been used as a visualization tool for solar data from different instruments \citep{Jhelioviewer_2017}. This research has made use of PyThea v0.7.4, an open-source and free Python package to reconstruct the 3D structure of CMEs and shock waves (Zenodo: \url{https://doi.org/10.5281/zenodo.5713659}). We also used GCS in Python v0.2.2 (Zenodo: \url{https://zenodo.org/badge/latestdoi/297350666}), which is a python 3 implementation of the Graduated Cylindrical Shell model based on the existing IDL implementation in SolarSoft. Other python(v3.9) libraries used for this work are: astropy (\cite{astropy2018AJ}),
 matplotlib
(\cite{matplotlib2007}), numpy (\cite{numpy2011}), pandas
(\cite{pandas2020}), scipy (\cite{scipy2020}), SunPy (\cite{sunpy2020ApJ}). The SolarSoft \footnote{\url{https://soho.nascom.nasa.gov/solarsoft/}} package in IDL(v8.6.0) has been used in this paper. Also the WISPR package on Solarsoft has been widely used.

\newpage

\end{document}